\documentclass{article}
\usepackage[affil-it]{authblk}
\usepackage[symbol]{footmisc}
\usepackage{todonotes}

\usepackage[scr=boondoxo]{mathalfa} 

\usepackage[utf8]{inputenc}
\usepackage[margin=2cm, letterpaper]{geometry}
\usepackage{braket,amsthm}
\usepackage{mathtools,amsfonts}
\usepackage{enumitem}
\usepackage{color,xcolor}
\usepackage{todonotes}
\usepackage{xfrac}

\usepackage{dsfont}
\definecolor{color1}{HTML}{FFB284}
\definecolor{color2}{HTML}{C6C09C}
\definecolor{color3}{HTML}{E79796}
\definecolor{color4}{HTML}{FCC98B}

\theoremstyle{definition}
\newtheorem{thm}{Theorem}[section]
\newtheorem{remark}[thm]{Remark}

\usepackage{hyperref} 
\hypersetup{
  colorlinks=true,
  linkcolor=color1,
  filecolor=color1,
  urlcolor=color1,
  citecolor=color1
}

\usepackage{tikz}
\usetikzlibrary{backgrounds,decorations.pathreplacing}
\tikzset{
operator1/.style = {draw, text=black, draw=color1,line width=1pt, fill=color1!10,minimum width=1cm,minimum size=1.5em},
operator2/.style = {draw, text=black, draw=color2,line width=1pt, fill=color2!10,minimum width=1cm,minimum size=1.5em},
operator3/.style = {draw, text=black, draw=color3,line width=1pt, fill=color3!10,minimum width=1cm,minimum size=1.5em},
brace0/.style = {decorate,decoration={brace,amplitude=5pt},black!50}}

\usepackage{subcaption}

\usepackage{pifont}	

\usepackage{bbm} 
\usepackage{colonequals} 
\usepackage{qcircuit} 
\newcommand{\qvdots}{\raisebox{0.3em}{\ensuremath{\vdots}}} 

\newcommand{\lb}{\left} 
\newcommand{\rb}{\right} 
\newcommand{\nn}{\nonumber} 
\newcommand{\qd}{\qquad} 
\newcommand{\qqd}{\qquad\qquad} 
\newcommand{\tx}{\text} 
\newcommand{\bd}{\boldsymbol} 
\newcommand{\mc}{\mathcal} 
\newcommand{\Or}{\mathcal{O}} 
\newcommand{\bb}{\mathbb} 
\newcommand{\id}{\mathbbm{1}} 
\newcommand{\ce}{\colonequals} 
\newcommand{\imp}{\implies} 
\newcommand{\tp}{\otimes} 
\newcommand{\ct}{\dagger} 
\newcommand{\infi}{\infty} 
\newcommand{\Bk}{\Braket} 

\newcommand{\iu}{\mathrm{i}} 

\DeclareMathOperator{\Var}{Var} 

\newcommand{\eqn}[1]{\begin{equation}#1\end{equation}} 
\newcommand{\eqns}[1]{\begin{equation*}#1\end{equation*}} 
\newcommand{\aln}[1]{\begin{align}#1\end{align}} 
\newcommand{\alns}[1]{\begin{align*}#1\end{align*}} 
\newcommand{\pmx}[1]{\begin{pmatrix}#1\end{pmatrix}} 

\newcommand{\Abs}[1]{\left|{#1}\right|} 
\newcommand{\Kb}[2]{\Ket{#1}\Bra{#2}} 
\newcommand{\opno}[1]{\left\|{#1}\right\|_{\text{op}}} 

\makeatletter
	\newcommand{\vast}{\bBigg@{3}}
	\newcommand{\Vast}{\bBigg@{4}}
\makeatother

\renewcommand{\Re}{\operatorname{Re}}

\AtBeginDocument{%
	\renewcommand{\Pr}{\operatorname{Pr}}
}

\title{Quantum advantage without exponential concentration: Trainable kernels for symmetry-structured data}

\author[1]{Laura J. Henderson}
\author[2]{Kerstin Beer\thanks{Correspondence to: kerstin.beer@mq.edu.au}}
\author[3]{Salini Karuvade}
\author[1]{Riddhi Gupta}
\author[1]{Angela White}
\author[1]{Sally Shrapnel}

\affil[1]{School of Mathematics and Physics, The University of Queensland, Brisbane, Australia}
\affil[2]{School of Mathematical and Physical Sciences, Macquarie University, Sydney, Australia}
\affil[3]{School of Physics, The University of Sydney, Sydney, Australia}

\date{}

\begin{document}
\maketitle

\begin{abstract}

Quantum kernel methods promise enhanced expressivity for learning structured data, but their usefulness has been limited by kernel concentration and barren plateaus. Both effects are mathematically equivalent and suppress trainability. We analytically prove that covariant quantum kernels tailored to datasets with group symmetries avoid exponential concentration, ensuring stable variance and guaranteed trainability independent of system size. Our results extend beyond prior two-coset constructions to arbitrary coset families, broadening the scope of problems where quantum kernels can achieve advantage. We further derive explicit bounds under coherent noise models—including unitary errors in fiducial state preparation, imperfect unitary representations, and perturbations in group element selection—and show through numerical simulations that the kernel variance remains finite and robust, even under substantial noise. These findings establish a family of quantum learning models that are simultaneously trainable, resilient to coherent noise, and linked to classically hard problems, positioning group-symmetric quantum kernels as a promising foundation for near-term and scalable quantum machine learning.
\end{abstract}

\section{Introduction}
The interdisciplinary field at the intersection of quantum computing and machine learning - quantum machine learning (QML) - seeks to harness quantum mechanical resources to perform learning algorithms. Broadly, QML research falls into three main categories \cite{Aimeur2006}: (i) quantum algorithms that accelerate classical ML tasks \cite{Paparo2014, Schuld2014, Wiebe2016}; (ii) classical ML techniques used to analyze and characterize quantum systems \cite{Lovett2013, Carleo2017, Tiersch2015}; and (iii) quantum devices employed to learn from quantum data \cite{Alvarez2017, Amin2018, Beer2020}.

A powerful technique of classical  machine learning are kernel methods \cite{bach_information_2023}.  These map input data into high-dimensional feature spaces and evaluate inner products between data points there. The so-called kernel trick allows to learn complex patterns without the computational burden of directly working in high dimensions. Quantum kernel \cite{schuld_supervised_2021, jerbi_quantum_2023, deshpande_dynamic_2024, wood_kerr_2024, kairon_equivalence_2025} methods extend this idea by using quantum computers to evaluate these inner products in exponentially large Hilbert space: classical data is encoded into quantum states via quantum feature maps, and the overlap between these states is computed by the quantum device which offers enhanced expressivity. 

As quantum hardware is advancing, there is growing interest in identifying problems for which QML can offer a clear computational advantage over classical methods. One promising approach is to construct a trainable problem instance inspired by quantum algorithms. In the following case, this instance is derived from the discrete logarithm problem (DLP) and thus inherits its classical computational hardness. The authors of \cite{liu_rigorous_2021, glick_covariant_2024} introduce a novel class of quantum kernels tailored for datasets exhibiting group symmetries. These kernels use unitary representations of  the underlying group to construct feature maps that are covariant under  the group action. This enables the  algorithm to learn symmetries present in the data. The authors demonstrate, both theoretically and experimentally (on a 27-qubit superconducting quantum processor), that such covariant kernels can be used to effectively classify data defined on coset-structured tasks that are challenging for classical models unless the symmetry is explicitly encoded. This family of quantum kernels is especially notable because it extends the kernel studied in \cite{liu_rigorous_2021}, which was capable of learning the task inspired by the DLP.

Despite these promises, the trainability of QML methods remains a key challenge. For quantum kernel methods, a major obstacle is kernel concentration \cite{thanasilp_exponential_2022}, where kernel values concentrate exponentially and the classifier loses expressive power. Interestingly, kernel concentration is closely related to barren plateaus in variational quantum circuits \cite{mcclean_barren_2018, cerezo_cost_2021, wang_noise-induced_2021, cerezo_higher_2021, arrasmith_effect_2021, holmes_barren_2021, ortiz_marrero_entanglement-induced_2021, patti_entanglement_2021, pesah_absence_2021, arrasmith_equivalence_2022, sharma_trainability_2022}; the presence of one implies the other \cite{thanasilp_exponential_2022,kairon_equivalence_2025}. While various strategies have been proposed to mitigate barren plateaus \cite{verdon_learning_2019-1, grant_initialization_2019, cerezo_cost_2021, uvarov_barren_2021, volkoff_large_2021, bilkis_semi-agnostic_2023,  sannia_engineered_2023} an equally desirable long-term goal is to design QML approaches that inherently avoid kernel concentration and its associated trainability issues.
 
In this work we analytically prove that quantum kernel methods tailored to data with coset group structure do not exhibit barren plateaus, ensuring trainability regardless of system size. 
Furthermore, we show that model remain trainable in the presence of various sources of coherent noise, including   errors in fiducial state preparation, imperfect unitary representations, and perturbations in group element selection.
Our numerical simulations support these results, demonstrating stable kernel variance and effective classification across a range of parameters. Together, these findings characterize a family of quantum learning models that are provably trainable, resilient to coherent noise, and grounded in classically difficult problems, thereby establishing group-symmetric quantum kernels as a promising framework for both near-term and scalable quantum machine learning.

In Section~\ref{sec:methods}, we present the methods and theoretical tools underpinning our analysis. Section~\ref{sec:groupkernel} introduces covariant quantum kernels tailored for data with group structure, while Section~\ref{sec:quantumKernels} reviews the broader framework of quantum kernels. Section~\ref{sec:barrenPlateaus} discusses key trainability challenges, focusing on barren plateaus and exponential concentration effects.
Section~\ref{sec:results} contains our main results. In Section~\ref{sec:results_variance}, we derive analytical expressions for the variance of the coset quantum kernel. Section~\ref{sec:results_numerics} reports numerical simulations validating these results. Section~\ref{sec:results_noise} extends the analysis to noisy systems, examining the impact of errors in the fiducial state (Section~\ref{sec:results_noise_errorFiducial}), the unitary representation of group elements (Section~\ref{sec:results_noise_unitary}), and the selection of group elements (Section~\ref{sec:results_noise_selectionError}), as well as providing numerical simulations of these errors (Section~\ref{sec:results_noise_numerics}). 
Finally, Section~\ref{sec:discussion} discusses the implications of our findings, their limitations, and potential directions for future research.

\section{Methods}
\label{sec:methods}
In this section, we describe the techniques and theoretical tools underlying our approach. We begin with an overview of covariant quantum kernels designed for data with group structure, followed by a brief review of quantum kernel methods. We then discuss challenges in trainability, particularly the issue of barren plateaus and concentration phenomena in parameterized quantum circuits.

\subsection{Learning on data with group structure}
\label{sec:groupkernel}

In the following we introduce the class of quantum kernels presented in  \cite{glick_covariant_2024}, which are specifically designed to learn datasets with an underlying group structure.
Specifically, it is assumed that the data belongs to either of the two cosets obtained from a subgroup $S \subset \mc{G}$ of the group $\mc{G}$. The cosets are obtained from $S$ by the left or right action of the group elements $c_1,c_2\in \mc{G}$ on the elements of  $\mc{G}$.
Let us denote the dataset by $\mc{X}:=c_1S\cup c_2 S$.
The task is to assign a label $y\in\{-1,1\}$ to a datum $x\in\mc{X}$, where the classification rule, also referred to as the \textit{concept}~\cite{liu_rigorous_2021}, is determined by the coset to which $x$ belongs.
A special class of quantum kernels called the covariant kernels was introduced in~\cite{glick_covariant_2024} to learn the concept class for problems with coset structure. 
In particular, the concept class chosen in \cite{liu_rigorous_2021} was based on the discrete log problem (DLP) and it was shown to be a learning problem with the same classical hardness as the DLP itself.


We now discuss the above terms in detail and also extend the learning problem to the case of more than two cosets.
To embed the classical data $\mc{X}\subset\mc{G}$ into the quantum setting we define a unitary representation $D:\mc{G}\to U(2^N)$ of the group $\mc{G}$ that acts on $N$ qubits.
Given a homogeneous space induced by some subgroup $S$ of $\mc{G}$, we define $m\ge2$ distinct left cosets for the group elements $\bd{c}_{j}\in \mc{G}$, $1\le j\le m$ by
\begin{equation*}
       C_j = \bd{c}_{j} S, 
\end{equation*}
where $\{\bd{c}_{j}\}$ is drawn randomly, for instance, according to the Haar measure on $\mc{G}$. We also assume that the number of cosets $m$ is independent of $n$ and $N$.
The learning goal is to categorise data
	$\bd{x} \in \bigcup_{j} C_j = \mc{X}$
into the correct coset.

The algorithm introduced in~\cite{glick_covariant_2024} to solve the above learning problem requires a reference state $\Ket{\psi}$, which is invariant under the action of the subgroup $S$
\begin{equation*}
	D_{\bd{s}}\Ket{\psi} = \Ket{\psi} \tx{ for all } \bd{s}\in S
\end{equation*}
and  
\begin{equation*}
	D_{\bd{x}}\Ket{\psi} = D_{\bd{c}_{j}\bd{s}}\Ket{\psi} = D_{\bd{c}_{j}}D_{\bd{s}}\Ket{\psi} = D_{\bd{c}_{j}}\Ket{\psi}.
\end{equation*}
for all $\bd{x}\in\mc{X}$. 
By using this reference state, the classifier in \cite{glick_covariant_2024} only needs to be able to distinguish between the $m$ states
\begin{equation*}
	\rho_{j} = D_{\bd{c}_{j}}\Kb{\psi}{\psi}D_{\bd{c}_{j}}^{\ct}, \quad 1\le j \le m.
\end{equation*}

\begin{figure}[t!]
    \centering
    \begin{subfigure}[b]{0.3\textwidth}
        \centering
\includegraphics[width=0.44\textwidth]{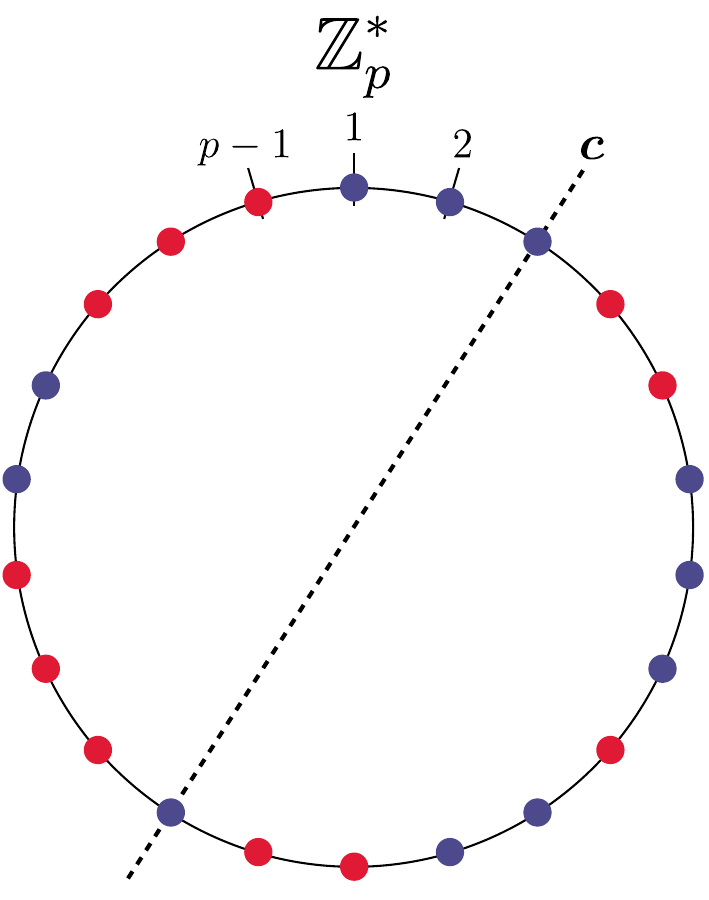}
        \hspace{1em}
\includegraphics[width=0.44\textwidth]{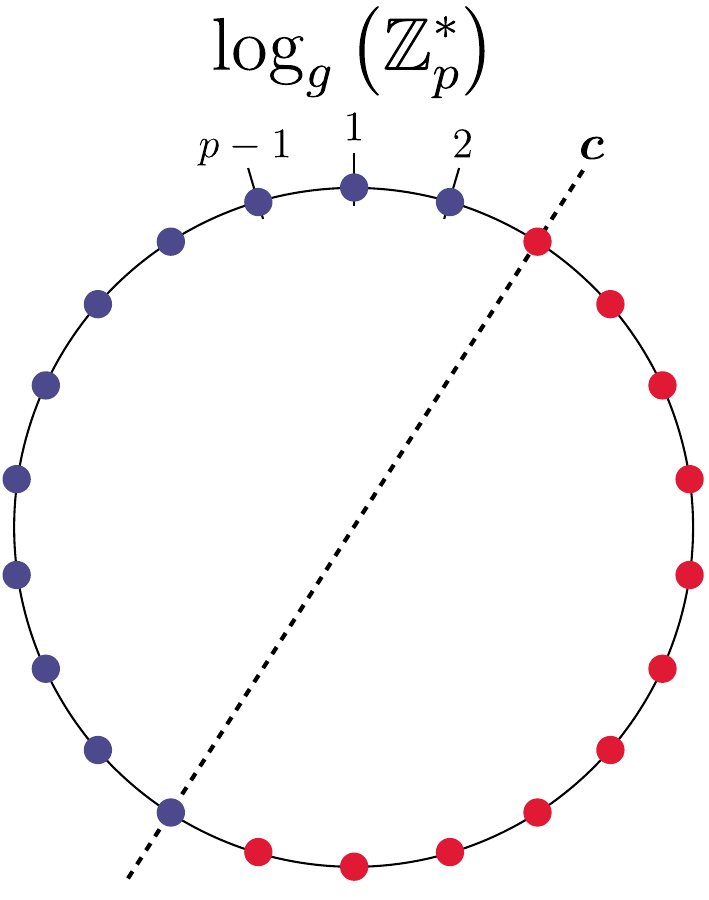}  
        \caption{ } 
        \label{fig:subfig_a}
    \end{subfigure}
    \hfill
    \begin{subfigure}[b]{0.3\textwidth}
        \centering
        \mbox{
		\Qcircuit @C=0.7em @R=0.5em {
	 		\Ket{0} & & \multigate{3}{V} & \multigate{3}{D_{\bd{x}}} & \qw & \multigate{3}{D_{\bd{x}'}^{\ct}} & \multigate{3}{V^{\ct}} & \meter & \cw \\
			\Ket{0} & & \ghost{V} & \ghost{D_{\bd{x}}} & \qw & \ghost{D_{\bd{x}'}^{\ct}} & \ghost{V^{\ct}} & \meter & \cw \\
			\qvdots & & \nghost{V}  & \nghost{D_{\bd{x}}} & & \nghost{D_{\bd{x}'}^{\ct}} & \nghost{V^{\ct}} & \qvdots \\
			\Ket{0} & & \ghost{V} & \ghost{D_{\bd{x}}} & \qw & \ghost{D_{\bd{x}'}^{\ct}} & \ghost{V^{\ct}} & \meter & \cw \\
			& & \dstick{\qqd U(\bd{x})} & & & \dstick{\qqd U^{\ct}(\bd{x}')}
			\gategroup{1}{3}{4}{4}{.7em}{--}
			\gategroup{1}{3}{4}{4}{1.4em}{_\}}
			\gategroup{1}{6}{4}{7}{.7em}{--}
			\gategroup{1}{6}{4}{7}{1.4em}{_\}}
		} 
	}  
\vspace{0.5cm}
        \caption{ } \label{fig:subfig_b}
    \end{subfigure}
    \hfill
    \begin{subfigure}[b]{0.3\textwidth}
    \centering
 \includegraphics[width=0.9\textwidth]{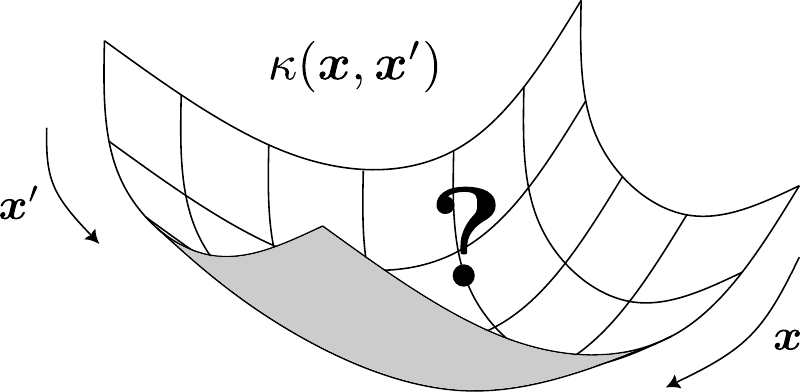} 
 \vspace{0.5cm}
        \caption{ } \label{fig:subfig_c}
    \end{subfigure}
            \caption{\textbf{Overview of the quantum kernel and concept class studied in this work.} (a) An illustration of the problem of learning the concept class, $\mc{C}$ \cite{liu_rigorous_2021}. In the $\bb{Z}_p^{*}$ group (where $p$ is prime) the data is scrambled (\textit{left}), but after taking the discrete log (DLOG), the data is easily separated by the concept $\bd{c}$ (\textit{right}). (b) The quantum circuit for calculating matrix elements of a covariant quantum kernel, $\kappa(\bd{x},\bd{x}')$. Each datum $\bd{x}\in\mc{X}$ is encoded in the unitary $D_{\bd{x}}$ and $V$ creates the fiducial state. This type of kernel can be used to learn the concept class (a) \cite{glick_covariant_2024}. (c) An illustration of the landscape of a quantum kernel, $\kappa(\bd{x},\bd{x}')$ as a function of it's inputs $\bd{x}$ and $\bd{x}'$. The landscape is described by the kernel variance, Equation~\eqref{eq:KernelVar}, and is a measure of the performance of the quantum kernel \cite{thanasilp_exponential_2022}. }
\end{figure}

\subsection{Quantum kernels}
\label{sec:quantumKernels}
Kernel methods are a powerful classical machine learning technique that enables learning in high-dimensional spaces via inner products, and quantum kernel methods extend this by using quantum computers to compute overlaps in exponentially large Hilbert spaces for greater expressivity. A \emph{quantum kernel} is a function 
\begin{equation*}
\kappa(x, x') = |\langle \phi(x) | \phi(x') \rangle|^2
\end{equation*}
capturing the similarity between inputs $x$  and  $x'$  in a quantum feature space. Here the classical data input points $ x, x' \in \mathcal{X}$ are mapped onto   quantum states $|\phi(x)\rangle, |\phi(x')\rangle$, respectively, in a Hilbert space $\mathcal{H}$, via the feature map
\begin{equation*}
x \mapsto |\phi(x)\rangle = U_\phi(x) |0\rangle.
\end{equation*}
Here $U_\phi(x)$ is a parameterized quantum circuit dependent on the input  $x$ , and $\ket{0} $ is the all-zero reference state.
Furthermore, the overlap between the two data points $x,x'$ is computed by evaluating the following kernel function on a quantum computer
\begin{equation*}
    \kappa(\bd{x},\bd{x}') = \Abs{\langle \phi(x')|\phi(x)\rangle}^2.
\end{equation*}
In the specific case of learning data with group structure discussed above, one can use the \textit{covariant} quantum kernel given in~\cite{glick_covariant_2024}
\begin{align*}
	\kappa(\bd{x},\bd{x}') &= \Abs{\Bk{0^{\tp N}|V^{\ct}D_{\bd{x}}^{\ct}D_{\bd{x}'}V|0^{\tp N}}}^2
	= \Abs{\Bk{\psi|D_{\bd{x}}^{\ct}D_{\bd{x}'}|\psi}}^2.
	%
\end{align*}
Figure~\ref{fig:subfig_b} illustrates the quantum circuit employed to evaluate this covariant kernel.



\subsection{Exponential concentration and barren plateaus}
\label{sec:barrenPlateaus}
As QML gains attention for its potential to outperform classical methods, a key challenge remains its limited trainability due to barren plateaus. For a model with loss function $l(\rho, O) $, defined as the variance of the expectation value of the observable $O$ over the quantum state $\rho$, we say the method reaches a \emph{barren plateau} when the variance of the loss over parameters $\boldsymbol{\theta}$ vanishes exponentially in the number of qubits $N$ as
\begin{equation*}
\operatorname{Var}_{\boldsymbol{\theta}}\big[ \ell_{\boldsymbol{\theta}}(\rho, O) \big] \in \mathcal{O}\bigl(1/b^{N}\bigr),
\end{equation*}
for some constant $b > 1 $ \cite{ragone_lie_2024}. 

Furthermore, the exponential concentration of a quantum kernel $\kappa(\bd{x},\bd{x}')$ in the number of qubits $N$ is defined as
\begin{align}
	\Pr\lb[\big|\kappa(\bd{x},\bd{x}')-\mu\big|\ge\delta\rb] &\le \frac{\beta}{\delta^2}, \quad \beta\in\Or(1/b^N) \label{eq:EC}\\
	\mu &= \bb{E}_{\bd{x},\bd{x}'}\big[\kappa(\bd{x},\bd{x}')\big] \nn
\end{align}
for some $b>1$ and where $\bb{E}_{\bd{x},\bd{x}'}\big[\kappa(\bd{x},\bd{x}')\big]$ is the expectation value of the kernel \cite{thanasilp_exponential_2022}. The authors of \cite{thanasilp_exponential_2022} show that if the kernel landscape (illustrated in Figure~\ref{fig:subfig_c}) becomes exponentially flat, meaning that if $\mu$ is exponentially small $(\mu\in \Or(1/b^N))$, then the resulting model (after a polynomial number of measurements) does not depend on the input data.

There is an equivalence between barren plateaus in variational quantum circuits and the exponential concentration of quantum machine learning kernels \cite{thanasilp_exponential_2022,kairon_equivalence_2025}, showing that the presence of one implies the other. 
Finally, we note that Chebyshev's inequality allows us to cast the existence of exponential concentration (equation~\eqref{eq:EC}) in terms of the variance of kernel values
\aln{
    \Var_{\bd{x},\bd{x}'}\big[\kappa(\bd{x},\bd{x}')\big] \in \Or(1/b'^N)
    \label{eq:KernelVar}
}
for some $b'>1$\cite{thanasilp_exponential_2022,kairon_equivalence_2025}. It is this definition of exponential concentration that we will use for the rest of the paper.

\section{Results}
\label{sec:results}
This section presents our main theoretical and numerical results. We derive variance expressions for covariant quantum kernels, analyze their robustness to errors coming from various sources of coherent noise, and identify conditions under which kernel concentration is avoided. Simulations support the theory across relevant regimes.

\subsection{Variance of the coset quantum kernel}
\label{sec:results_variance}
While such a kernel was already proposed in \cite{glick_covariant_2024} for the case of two cosets, we extend the definition here so that it can be applied to datasets containing more than two cosets. In practice, each kernel evaluation still compares states from only two cosets at a time, but when multiple cosets are present in the data, the resulting kernel matrix naturally includes entries for all pairwise comparisons, leading to a larger matrix. For now, we consider the idealized case without noise. The impact of noise will be addressed in Section~\ref{sec:noise}. By leveraging the properties of the fiducial state, the only kernel values are those that depend on the hidden group elements, $\bd{c}_{i}$:
\begin{align}
	\kappa(\bd{c}_{i}\bd{s}_a,\bd{c}_{i}\bd{s}_b) &= \Abs{\Bk{\psi|D_{\bd{c}_{i}}^{\ct} D_{\bd{c}_{i}}|\psi}}^2 = 1 \nn\\
	\kappa(\bd{c}_{i}\bd{s}_a,\bd{c}_{j}\bd{s}_b) &= \Abs{\Bk{\psi|D_{\bd{c}_{i}}^{\ct} D_{\bd{c}_{j}}|\psi}}^2 \equalscolon \alpha_{i,j}=\alpha_{j,i} < 1 \quad i\ne j\label{eqn:alpha}
\end{align}
for any $s_a,s_b \in S$.

Since each coset has the same number of elements, each kernel matrix will always have $mn^2$ occurrences of $\kappa=1$ and $2n^2$ occurrences of each $\kappa=\alpha_{i,j}$ (after accounting for the fact that $\alpha_{i,j}=\alpha_{j,i}$), where $n=|S|$ and $m\ge2$ the number of distinct left cosets. When considering the distribution of kernel values, we will exclude the $mn$ trivial kernel values that appear in the diagonal, $\kappa(\bd{c}_{i}\bd{s}_a,\bd{c}_{i}\bd{s}_a)=1$ \cite{thanasilp_exponential_2022}. Now the expectation value of the kernel value is
\begin{align*}
    \bb{E}_{\bd{x},\bd{x}'}\big[\kappa(\bd{x},\bd{x}')\big] &= \frac{1}{(mn)^2-mn}\lb[m(n^2-n)+2n^2\sum_{i=1}^{m-1}\sum_{j=i+1}^{m}\alpha_{i,j}\rb] \nn\\
    &= \frac{1}{m(mn-1)}\lb[m(n-1)+2n\sum_{i=1}^{m-1}\sum_{j=i+1}^{m}\alpha_{i,j}\rb]
\end{align*}
and the variance will be
\begin{align*}
    &\Var_{\bd{x},\bd{x}'}\big[\kappa(\bd{x},\bd{x}')\big] \nn\\
    &\qd = \frac{1}{(mn)^2-mn}\lb[m(n^2-n)\Big(1-\bb{E}_{\bd{x},\bd{x}'}\big[\kappa(\bd{x},\bd{x}')\big]\Big)^2 + 2n^2\sum_{i=1}^{m-1}\sum_{j=i+1}^{m}\Big(\alpha_{i,j}-\bb{E}_{\bd{x},\bd{x}'}\big[\kappa(\bd{x},\bd{x}')\big]\Big)^2\rb] \nn\\
    &\qd = \frac{1}{m^2(mn-1)^2}\vast[m(mn-1)\Bigg(m(n-1)+2n\sum_{i=1}^{m-1}\sum_{j=i+1}^{m}\alpha_{i,j}^2\Bigg) - \Bigg(m(n-1)+2n\sum_{i=1}^{m-1}\sum_{j=i+1}^{m}\alpha_{i,j}\Bigg)^2\vast].
\end{align*}

Since each $\alpha_{i,j}$ is smaller than 1, we note that neither the expected value nor the variance of the kernel is vanishing. This becomes clearer as we consider the large $N$ limit below.

In the limit of large $N$, $\alpha_{i,j}$ 
can be approximated by its expected value taken over all possible choices of $c_i,c_j$, where these two group elements are sampled randomly according to some chosen measure on $\mc{G}$.
Hence the expected value of $\alpha_{i,j}$ depends on both the measure chosen as well as on the unitary representation of $\mc{G}$.
However, as noted above, for any nontrivial choice of this measure we have $\alpha_{i,j}<1$ and hence the relevant quantities are nonvanishing.
To make our argument more concrete,
we consider the special case where $c_i,c_j$ are sampled according to Haar measure defined on the $N$-qubit unitary group.
For large $N$, we can reasonably approximate

\begin{equation*}
    \alpha_{i,j} \approx \frac{1}{2^N}
\end{equation*}
for all $i\ne j$ (see Appendix~\ref{appx:alpha} for details).
Then,
\begin{equation*}
	\bb{E}_{\bd{x},\bd{x}'}\big[\kappa(\bd{x},\bd{x}')] \approx \frac{1}{mn-1}\lb((n-1)+n(m-1)\frac{1}{2^N}\rb) \xrightarrow[N\to\infi]{} \frac{n-1}{mn-1}
\end{equation*}
and
\begin{equation}
    \Var_{\bd{x},\bd{x}'}\big[\kappa(\bd{x},\bd{x}')\big] \approx \frac{n(n-1)(m-1)}{(mn-1)^2}\lb(1-\frac{1}{2^N}\rb)^2 \xrightarrow[N\to\infi]{} \frac{n(n-1)(m-1)}{(mn-1)^2}.
    \label{eq:VarNoError}
\end{equation}

Finally, we note that in the case that $n\to\infi$ as $N\to\infi$, as is the case in the protocol proposed in \cite{glick_covariant_2024} on which we base our numerical simulations, then
\begin{equation*}
	\bb{E}_{\bd{x},\bd{x}'}\big[\kappa(\bd{x},\bd{x}')] \xrightarrow[N\to\infi]{} \frac{1}{m}
\end{equation*}
and
\begin{equation}
    \Var_{\bd{x},\bd{x}'}\big[\kappa(\bd{x},\bd{x}')\big] \xrightarrow[N\to\infi]{} \frac{m-1}{m^2}.
    \label{eq:VarNoError2}
\end{equation}

That is, even if the data is partitioned into Haar-randomly chosen cosets, there is so much structure in the problem that this group-symmetric learning problem is free of exponential concentration. 
This is desirable because it ensures that the kernel retains the ability to meaningfully distinguish between different input data points. If the variance of the kernel values were to vanish, all inputs would appear nearly identical in the quantum feature space, rendering the kernel uninformative for learning tasks. This situation is analogous to the barren plateau problem in variational algorithms, where gradients vanish and optimization becomes ineffective. 
 While this is an interesting theoretical result, it remains somewhat idealized. To move toward a more realistic setting, we introduce noise to the cosets in Section~\ref{sec:noise}. 
 Finally, we end this section with the following remark.

 \begin{remark}\label{rem:subset}
     We note that the same results will hold if $S$ is only a subset of a group and does not form a subgroup, provided that (i) a fiducial state can be found so that $s\Ket{\psi}=\Ket{\psi}$ for all $s\in S$ and (ii) each set $C_i=c_iS$ is disjoint so that if $x\in C_i$ then $x\notin C_j$ for all $j\ne i$.
 \end{remark}

\subsection{Numerical results}
\label{sec:results_numerics}
Our analytical results establish the theoretical foundation for the absence of barren plateaus in the considered setting. We now show that these findings are further supported by numerical simulations, which confirm the predicted behavior across a range of system sizes and parameters. 

\begin{figure}[h!]
	\centering
        \includegraphics[width=0.75\textwidth]{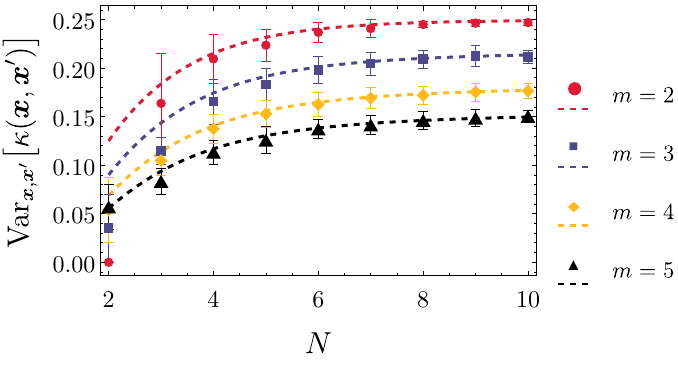}
	\caption{
		An average of 100 simulated values of variance in the kernel, $\Var_{\bd{x},\bd{z}}\big[\kappa(\bd{x},\bd{z})\big]$, (dots) compared to the theoretical variance (dashed lines) as a function of the number of qubits, $N$, for various number of disjoint sets, $m$, for perfect encoding. The kernel is constructed from training data, which consists of half of the data, randomly selected for each trial, such that each coset is represented at least once. Error bars represent one standard deviation.
	}
        \label{fig:SimNoErrors}
\end{figure}

We perform classical simulations of the protocol proposed in~\cite{glick_covariant_2024} for chain graph states of 2 to 10 qubits. We provide a review of this protocol in Appendix~\ref{appx:SpecificGroup}. 
The group $\mc{G}$ under consideration comprises arbitrary single-qubit rotations on all $N$ qubits and the subgroup $S_{\rm graph}$ is generated by the graph stabilizers of the form given in equation~\eqref{eq:Sgraph}.
However, it is crucial to note here that
rather than working with the full subgroup with $|S_{\rm graph}|=2^N$, we work with a subset of the same.
That is, we
take $S$ to be the set of generators of $S_{\rm graph}$
so that $|S|=n=N$ (see equation~\eqref{eq:ChainGenerators}). 
This guarantees that our sets $C_i=c_iS$ are disjoint and the size of the dataset $\mc{X}$ does not scale unfavorably with $N$.
Following Remark~\ref{rem:subset},
our analytical results still hold for this modified setting, while at the same time 
keeping the computational cost down. The simulated values closely match the theoretical predictions for approximately \( N > 6 \), confirming the expected behavior in the asymptotic regime: Figure~\ref{fig:SimNoErrors} shows that if neither errors nor kernel alignment are included, this problem reduces to finding the distance between two random states, and the variance (through basic simulation) appears to asymptote to $(m-1)/m^2\in \Or(1)$ (equation~\eqref{eq:VarNoError2}) with increasing $N$, which is consistent with our theoretical results.

\begin{figure}[ht]
	\begin{center}
    \includegraphics[width=\textwidth]{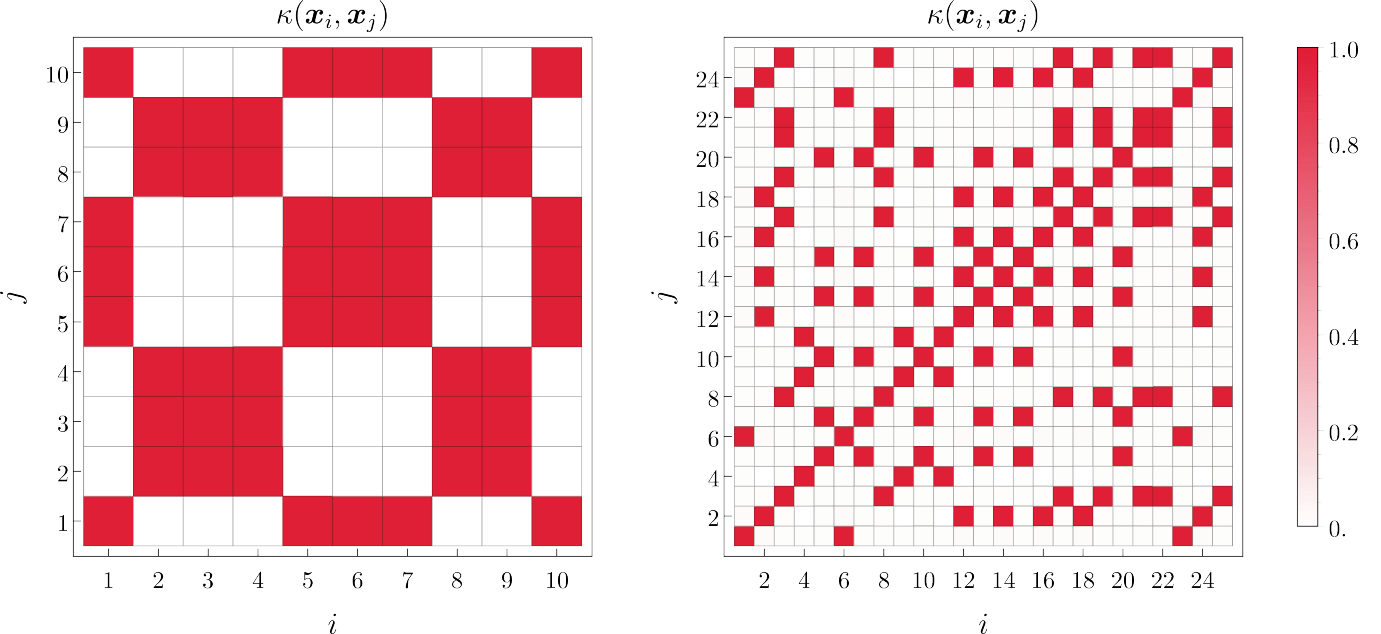}
	\caption{
            A heat map showing the kernel values for one simulation of $N=10$ qubits for (\textit{left}) $m=2$ and (\textit{right}) $m=5$ disjoint subsets. In both cases, there are many off-diagonal kernel entries where $\kappa(\bd{x}_i,\bd{x}_j)=1$ (shown in red) and do not approach the identity matrix. Note that the white entries are not exactly zero, but correspond to very small values $(<5.3\times10^{-3})$.
		}
    \label{fig:HeatMap}
	\end{center}
\end{figure}

In Figure~\ref{fig:HeatMap}, we plot a heat map of the kernel matrix of two instances of our numerical simulations, where we show in both cases there remain many non-diagonal kernel entries where $\kappa=1$. This highlights our main theoretical result, where the absence of kernel concentration results from the fact that the number $\lb(\frac{n-1}{mn-1}\rb)$ of off-diagonal kernel entries with $\kappa=1$ approaches a constant $(1/m)$ as $n$ becomes large.

A potential concern arises from the construction of the kernel using only the training data, which is subsequently applied to the testing data. If the dataset $\mc{X} $ is randomly partitioned, the training subset may not be uniformly distributed across all cosets. As a result, the variance will deviate from the expected $(m - 1)/m^2$. However, our numerical simulations confirm that, on average, the variance remains consistent with theoretical expectations across random data splits, where training data consist of half the total data such that at least one element from each coset appears in the training data (Figure~\ref{fig:SimNoErrors}).

\subsection{Extension to Noisy Systems}\label{sec:noise}
\label{sec:results_noise}
Simulating noise allows us to evaluate the robustness and practical viability of quantum algorithms. In this section, we focus on coherent errors as a first step to showing robustness under realistic operating conditions.

We assume that any errors will take the form
\begin{align*}
    	\kappa(\bd{x}_{a,i},\bd{x}_{b,i}) &= 1 - \gamma_{ab,i} \nn\\
	\kappa(\bd{x}_{a,i},\bd{x}_{b,j}) &= \alpha + \delta_{ab,ij},
\end{align*}
where $x_{a,i}:=c_as_i$, $\gamma_{ab,i}>0$ and $\delta_{ab,ij}\in\bb{R}$ and we assume that $\alpha_{i,j}=\alpha$ for all $1\le i,j\le m$ with $i\ne j$. For notational convenience, we vectorize \( \bd{\gamma} \in \mathbb{R}^{mn(n-1)} \) and \( \bd{\delta} \in \mathbb{R}^{m(m-1)n^2} \).

This yields an expected kernel value of
\begin{equation*}
	\bb{E}_{\bd{x},\bd{x'}}\big[\kappa(\bd{x},\bd{x'})\big] = \frac{n-1}{mn-1}\big(1-\bb{E}[\bd{\gamma}]\big) + \frac{n(m-1)}{mn-1}\big(\alpha-\bb{E}[\bd{\delta}]\big)
\end{equation*}
where $\bb{E}[\bd{\gamma}]$ and $\bb{E}[\bd{\delta}]$ are the expected values of the $\gamma_{ab,i}$'s and the $\delta_{ab,ij}$'s, respectively, 
and a variance of
\begin{equation*}
    \Var_{\bd{x},\bd{x}'}\big[\kappa(\bd{x},\bd{x'})\big] = \frac{n(m-1)(n-1)}{(mn-1)^2} \Big(\big(1-\bb{E}[\bd{\gamma}]\big)-\big(\alpha+\bb{E}[\bd{\delta}]\big)\Big)^2 +\frac{(n-1)\Var[\bd{\gamma}]+n(m-1)\Var[\bd{\delta}]}{mn-1}
    \label{eqn:ErrorVarMoreCoset}
\end{equation*}
where $\Var[\bd{\gamma}]$ and $\Var[\bd{\delta}]$ are the variances in the $\gamma_{ab,i}$'s and the $\delta_{ab,ij}$'s, respectively.

With that we establish a lower bound on the variance, showing that as long as \( 1 - \mathbb{E}[\bd{\gamma}] \) remains sufficiently large, the kernel does not concentrate. Intuitively, this is expected: as long as \( 1 - \gamma \) remains, on average, well-separated from \( \alpha + \delta \), the variance remains non-negligible, preventing concentration of the kernel. 

In the following sections, we examine how different sources of coherent noise affect the kernel values, focusing on errors in the fiducial state, the unitary representation of the group elements, and the group element selection process. In all three cases, we assume that the evaluated kernel values are still the result of the overlap of pure states, $\kappa(\bd{x},\bd{x}')=|\braket{\tilde{\phi}_{\bd{x}}|\tilde{\phi}_{\bd{x}'}}|^2$ and therefore any noise will map pure states to pure states. More specifically, given that the initial state $\ket{0^{\tp N}}\in\mc{H}$, we will assume that any operator $T:\mc{H}\to\mc{H}^*$ (where $\mc{H}^*$ is the dual to $\mc{H}$) that acts on $\ket{0^{\tp N}}$, including the operators that describe coherent noise, is a bounded linear operator, $T\in B(\mc{H})$. We use the operator norm (Appendix~\ref{appx:OpeartorNorm}) as a metric on $B(\mc{H})$.

\subsubsection{Errors in the Fiducial State}
\label{sec:results_noise_errorFiducial}
Suppose there is an implementation error in the operator \( V \), which prepares the fiducial state \( \ket{\psi} \) from the all-zero state \( \ket{0}^{\otimes N} \). That is, there exists bounded linear operators $W_k \in B(\mc{H})$ is  such that
\begin{equation*}
    \| V - W_k \|_{\tx{op}} \leq \epsilon.
\end{equation*}
Then, for any indices \( k, \ell \) and group elements strings \( \bd{c}_i, \bd{c}_j, \bd{s}_a, \bd{s}_b \), it follows that (to order $\epsilon^2$)
\begin{align*}
\Abs{\Bk{0^{\tp N}|W_k^{\ct}D_{\bd{s}_a}^{\ct}D_{\bd{c}_i}^{\ct}D_{\bd{c}_i}D_{\bd{s}_b}W_{\ell}|0^{\tp N}}}^2 &\ge 1 - 4\epsilon + 2\epsilon^2
\\
\alpha - 4\sqrt{\alpha}\epsilon + 2(2-\sqrt{\alpha})\epsilon^2 \le \Abs{\Bk{0^{\tp N}|W_k^{\ct}D_{\bd{s}_a}^{\ct}D_{\bd{c}_i}^{\ct}D_{\bd{c}_j}D_{\bd{s}_b}W_{\ell}|0^{\tp N}}}^2 & \le \alpha + 4\sqrt{\alpha}\epsilon + 2(2+\sqrt{\alpha})\epsilon^2.
    \label{eq:PsiErrorDifferent}
\end{align*}
Appendix~\ref{appx:errorFiducial} contains the complete derivation and Appendix~\ref{appx:noisy1} discusses the simulations of the errors.

\subsubsection{Error in the unitary representation of the elements of the group}
\label{sec:results_noise_unitary}
Now, we assume that there is an error in the unitary operator $D_{\bd{x}_{a,i}}$ such that
\begin{equation*}
    	\opno{D_{\bd{x}_{a,i}}-A_{a,i}} = \opno{D_{\bd{s}_a}D_{{c}_i}-A_{a,i}}\le \epsilon
\end{equation*}
where $A_{a,i}\in B(\mc{H})$.

We can show
\begin{align*}
\Abs{\Bk{\psi|A_{a,i}^{\ct}A_{b,i}|\psi}}^2 &\ge 1 - 4\epsilon + 2\epsilon^2
\\
\alpha - 4\sqrt{\alpha}\epsilon + 2(2-\sqrt{\alpha})\epsilon^2 \le \Abs{\Bk{\psi|A_{a,i}^{\ct}A_{b,j}|\psi}}^2 & \le \alpha + 4\sqrt{\alpha}\epsilon + 2(2+\sqrt{\alpha})\epsilon^2.
\end{align*}

The derivation of the bound is very similar to that for errors in the fiducial state and can be found in the Appendix~\ref{appx:unitaryRep}.

\subsubsection{Error in the selection of the group elements }
\label{sec:results_noise_selectionError}

Assume there exists a random perturbation satisfying
\begin{equation*}
    d(\bd{e}_{a,i}, \id) := \| D_{\bd{e}_{a,i}} - \id \|_{\tx{op}} < \epsilon,
\end{equation*}
such that the unitary operator \( D_{\bd{x}_{a,i}} \) admits the decomposition
\begin{equation*}
    D_{\bd{x}_{a,i}} = D_{\bd{e}_{a,i}} D_{\bd{q}_{a,i}} = D_{\bd{e}_{a,i}} D_{\bd{c}_i} D_{\bd{s}_a},
\end{equation*}
as described in~\cite{glick_covariant_2024}.

We prove to order $\epsilon^2$ that
\begin{align*}
  	&\Abs{\Bk{\psi|D_{\bd{c}_{\pm}}^{\ct}D_{\bd{e}_i}^{\ct}D_{\bd{e}_{j}}D_{\bd{c}_{\pm}}|\psi}}^2 \ge 1-\epsilon^2 \nn\\
    \alpha-4\sqrt{\alpha}\epsilon+4\epsilon^2\le &\Abs{\Bk{\psi|D_{\bd{c}_{\pm}}^{\ct}D_{\bd{e}_k}^{\ct}D_{\bd{e}_{\ell}}D_{\bd{c}_{\mp}}|\psi}}^2 \le \alpha+4\sqrt{\alpha}\epsilon+4\epsilon^2.
\end{align*}
See Appendix~\ref{appx:selectionError} for the derivation and Appendix~\ref{appx:noisy} for a discussion on the simulation of these errors.

\subsubsection{Numerical simulation of errors}
 \label{sec:results_noise_numerics}
\begin{figure}[ht]
	\begin{center}
    \includegraphics[width=\textwidth]{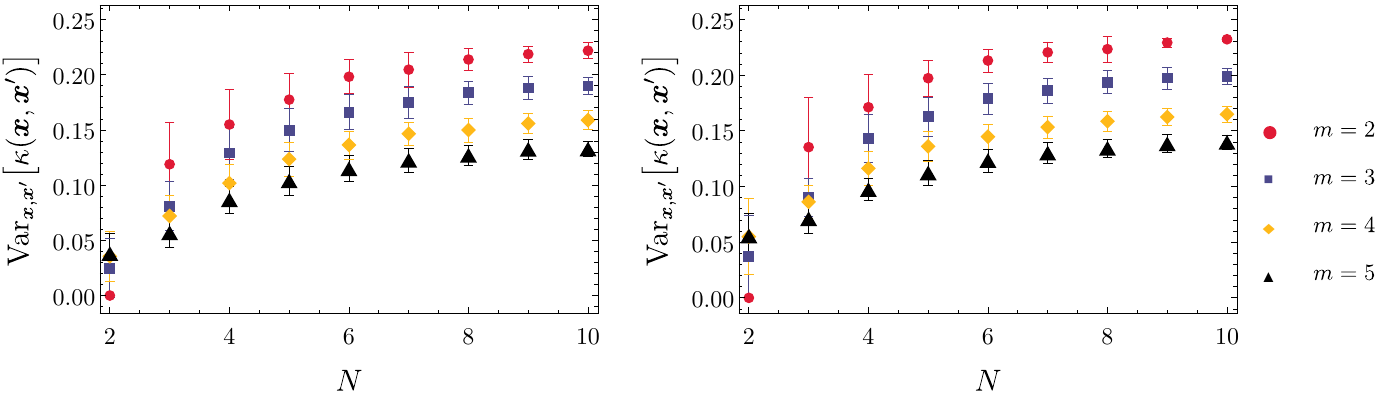}
	\caption{
            An average of 100 simulated values of variance in the kernel with errors, $\Var_{\bd{x},\bd{z}}\big[\kappa(\bd{x},\bd{z})\big]$, as a function of the number of qubits, $N$, for various number of disjoint sets, $m$, with imperfect kernel evaluation. The \textit{left} figure simulates an error in the fiducial state with $\opno{V-W_k}\le0.9$ and the \textit{right} figure simulates an error of the group elements with a perturbation of $\opno{D_{\bd{e}_{a,i}}-\id}\le0.9$. The kernel is constructed from training data, which consists of half of the data, randomly selected for each trial, such that each coset is represented at least once. In both cases, the variance still approaches a positive constant with increasing number of qubits, $\mc{O}(1)$, showing the absence of kernel concentration. Error bars represent one standard deviation.
		}
    \label{fig:SimErrors}
	\end{center}
\end{figure}

The theoretical errors calculated in the previous subsections represent the worst-case scenarios. In order to explore a slightly more experimentally relevant scenario, we numerically simulate errors in the evaluation of the quantum kernel, which are shown in Figure~\ref{fig:SimErrors}. Once again following the protocol proposed in~\cite{glick_covariant_2024} for chain graph states of 2 to 10 qubits, we simulate errors in the fiducial state (left plot) and perturbations in the group elements (right plot) with errors bounded by 0.9 in the operator norm (see Appendices~\ref{appx:noisy1} and~\ref{appx:noisy} for details). We find that although the error is large enough for the variance to show visible deviations from the ideal value, the kernel does not concentrate; its variance approaches a constant value as the number of qubits $(N)$ increases. Although this constant value is less than the ideal theoretical value given by Equation~\eqref{eq:VarNoError2}, it is much larger than the value given by the worst-case theoretical values above. We also find that errors in the fiducial state lead to a larger reduction of variance than perturbations in the group elements even though the allowed range of errors is the same in both cases. This is in line with the theoretical calculations, where the latter case results in a smaller error in the $\kappa(x_{\bd{a,i}},x_{\bd{b,i}})$ kernel elements. Finally we note, that as we do not simulate learning with these kernels, only the distribution of their entries, we do not comment on the effect of the simulated noise on their learning performance.

\section{Discussion}
\label{sec:discussion}

Our results highlight the power of quantum kernel methods in learning data with an underlying group structure, where classical approaches may struggle. We analytically demonstrated that the associated quantum kernel does not suffer from barren plateau phenomena. This stands in contrast to many variational quantum algorithms, where exponentially vanishing gradients can severely limit translatability as system size increases.

In addition, we rigorously analyzed the impact of noise on our method. We found that this quantum kernel construction is remarkably robust to common types of coherent noise—both analytically and through extensive numerical simulations. This robustness further strengthens the case for implementing our approach on near-term noisy intermediate-scale quantum (NISQ) devices. Unlike many quantum algorithms whose performance degrades significantly under realistic noise models, our method continues to perform reliably, demonstrating both stability and resilience. To better understand the robustness of the method, we examine how different noise sources influence the kernel values. These include errors in the preparation of the fiducial state, inexact unitary implementations, and deviations in the selection of group elements. 

An additional strength of our kernel family lies in its connection to earlier constructions. It generalizes a kernel previously shown to learn the concept class $\mathcal{C}$, which was based on the discrete logarithm problem (DLP). Since $\mathcal{C}$ inherits the classical hardness of the DLP, this places our construction in a particularly interesting category: quantum models that can efficiently learn tasks tied to classically intractable problems.

Together, these results position our quantum kernel method as a theoretically sound, resistant to coherent noise, and practically viable approach to learning structured data with quantum-enhanced models.


\textbf{Acknowledgments:} L.J.H., K.B., S.K., A.W. and S.S. are supported by the ARC Centre of Excellence for Engineered Quantum Systems (CE17010000) with K.B., S.K., and A.W. additionally supported via Deborah Jin Fellowships. R.S.G. is supported by the University of Queensland's Queensland Digital Health Center via funding from UQ’s Health Research Accelerator (HERA) initiative. 
Authors thank Rishi Goel, Jovana Gojkovic, and Connor Van Rossum for useful discussions.
\\
\textbf{Competing interests:} All authors declare no financial or non-financial competing interests.\\
\textbf{Data and code availability:} All scripts and data are available upon reasonable request and will be made publicly available at time of publishing.\\

\bibliographystyle{ieeetr}
\bibliography{bib.bib}


\appendix

\newpage
\section{Appendix}

\subsection{Discussion of the $\alpha_{i,j}$}
\label{appx:alpha}

In the following we discuss the value of $\alpha_{i,j}$, the overlap between two random pure states defined in Equation~\eqref{eqn:alpha} in Section~\ref{sec:results_variance}, where we derive the variance expressions for covariant quantum kernels.

For that, let $\ket{0}$ state be represented by the column vector $(1,0,\dots,0)^T$. Any vector distributed according to Haar measure can be written as $\ket{\phi} := U\ket{0}$ and $\ket{\psi} := V\ket{0}$ for some Haar-randomly sampled unitaries $U,V$. Now 
\begin{equation*}
    \bb{E}_{\phi,\varphi \sim \bb{C}^{2^N}}  = \Big[\Abs{\Bk{\phi|\varphi}}^2\Big] = \bb{E}_{U,V \sim \bb{C}^{2^N\times 2^N}}  = \Big[\Abs{\Bk{0|U^\dagger V|0}}^2\Big].
\end{equation*}
Note that Haar measure is invariant under left and right multiplication, therefore $W:= U^\dagger V$ is yet another Haar random unitary. We can now rewrite the overlap as
\begin{equation*}
     \bb{E}_{\phi,\varphi \sim \bb{C}^{2^N}} = \bb{E}_{W \sim \bb{C}^{2^N\times 2^N}}
    \Abs{W_{11}^2} = \int_0^1 (2^N-1)w(1-w)^{2^N-2}{\rm d}w = \frac{1}{2^N}, 
\end{equation*}
since the entry $|W_{11}|^2$ of a Haar-random unitary matrix \cite{petz_asymptotics_2004} $W \in \mathbb{C}^{d \times d}$ for $d:=2^N$ is distributed according to a Beta distribution:
\begin{equation*}
   |W_{11}|^2 \sim \mathrm{Beta}(1, d - 1) .
\end{equation*}
This yields the probability density function
\begin{equation*}
P(w) = (d - 1)(1 - w)^{d - 2}, \quad w \in [0, 1]\,.
\end{equation*}
This result follows from the fact that the columns of a Haar-random unitary are uniformly distributed unit vectors in $\mathbb{C}^d$, and the squared moduli of their components follow a Dirichlet distribution. The marginal distribution of a single component squared modulus is therefore Beta-distributed.

The expected value of a beta distribution is 
\begin{equation*}
\mathbb{E}[X] = \frac{\alpha}{\alpha + \beta}.
\end{equation*}
With parameters $\alpha = 1$ and $\beta = d - 1$, it is reasonable to approximate, in the limit of a large number of qubits, that
\begin{equation*}
    \alpha_{i,j} \approx \frac{1}{2^N}
\end{equation*}
for all $i \neq j$.

\subsection{Classifying cosets of $\tx{SU}(2)^{\tp N}$}
\label{appx:SpecificGroup}

In this section we briefly review the protocol proposed in~\cite{glick_covariant_2024} which we use to generate the numerical simulation shown in Figures~\ref{fig:SimNoErrors} and~\ref{fig:SimErrors}.

The chosen group is $\mc{G}=\tx{SU}(2)^{\tp N}$, which is the group of single qubit rotations on $N$ qubits. The natural unitary representation is a rotation $D(\theta_1,\theta_2,\theta_3)$ on each qubit where $\theta_1$, $\theta_2$, and $\theta_3$ are Euler angles
\eqns{
    D(\theta_1,\theta_2,\theta_3) = \exp\lb(-\frac{\iu\theta_1}{2}X\rb) \exp\lb(-\frac{\iu\theta_2}{2}Z\rb) \exp\lb(-\frac{\iu\theta_3}{2}X\rb) = R_x(\theta_1)R_z(\theta_2)R_x(\theta_3).
}

The choice of subgroup is a Pauli stablizer graph state on $N$ qubits. For a graph $G=(E,V)$, the graph-stabilizer is
\eqn{
    S_{\tx{graph}}=\lb\langle \Bigg\{X_j\bigotimes_{k:(j,k)\in E}Z_k\Bigg\}_{j\in V}\rb\rangle.
    \label{eq:Sgraph}
}
The graph state of the same graph $G=(E,V)$ is
\eqns{
    \Ket{G}=\prod_{(j,k)\in E} \tx{CZ}_{(j,k)} \Ket{+}^{\tp V} 
}
where $\tx{CZ}_{(j,k)}$ is the controlled-Z interaction between qubits $j$ and $k$. The state $\Ket{G}$ is the simultaneous $(+1)$-eigenvalue eigenvector of all of the elements of $S_{\tx{graph}}$:
\eqns{
    s\Ket{G} = \Ket{G} \quad \tx{for all } s\in S_{\tx{graph}}.
}
Therefore, $\Ket{G}$ matches the properties we require for a fiducial state.

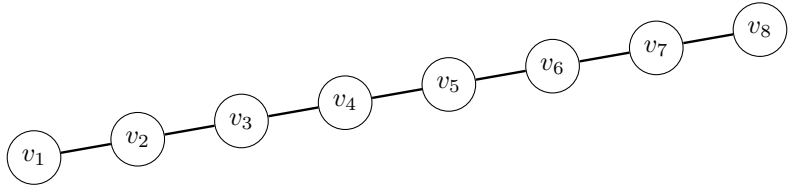
\begin{figure*}[h!]
	\begin{center}
		
		\begin{tikzpicture}[scale=1.4, rotate=10]
        \tikzstyle{place}=[circle,draw,minimum size=6mm]
			\node[place ] at (0,0) (a1) {$v_1$};
			\node[place ]   at (1,0) (a2)  {$v_2$}; 
			\node[place ]   at (2,0) (a3) {$v_3$};
			\node[place ] at (3,0) (a4) {$v_4$};
			\node[place ]  at (4,0) (a5) {$v_5$};
			\node[place ] at (5,0) (a6) {$v_6$};
			\node[place ] at (6,0) (a7) {$v_7$};
			\node[place ] at (7,0) (a8) {$v_8$};
			\draw[line width=0.9pt] (a1) -- (a2) -- (a3) -- (a4) -- (a5) -- (a6) -- (a7) -- (a8);
		\end{tikzpicture}
		\caption{A chain graph of 8 vertices.}
		\label{line}
	\end{center}
\end{figure*}

For simplicity, we choose the chain graph (Figure~\ref{line}), so our subgroup is the graph-stabilizer group whose generators are given by
\eqn{
   S:= \big\{X_1Z_2\id_3\dotsb\id_N, Z_1X_2Z_3\dotsb\id_N, \id_1Z_2X_3Z_4\dotsb\id_N, \dotsc, \id_1\id_2\id_3\dotsb Z_{N-1}X_N\big\}.
    \label{eq:ChainGenerators}
}

\begin{figure}[ht]
	\begin{center}
	\makebox[\textwidth]{
		\Qcircuit @C=0.7em @R=0.7em {
	 		\Ket{0} & & \gate{R_y(\frac{\pi}{2})} & \ctrl{1} & \qw & \qw & \gate{R_x(\theta_{1,1})} & \gate{R_z(\theta_{1,2})} & \gate{R_x(\theta_{1,3})} & \qw & \gate{R_x^{\ct}(\theta_{1,3}')} & \gate{R_z^{\ct}(\theta_{1,2}')} & \gate{R_x^{\ct}(\theta_{1,1}')} & \qw & \ctrl{1} & \qw & \gate{R_y^{\ct}(\frac{\pi}{2})} & \meter & \cw \\
			\Ket{0} & & \gate{R_y(\frac{\pi}{2})} & \ctrl{-1} & \ctrl{1} & \qw & \gate{R_x(\theta_{2,1})} & \gate{R_z(\theta_{2,2})} & \gate{R_x(\theta_{2,3})} & \qw & \gate{R_x^{\ct}(\theta_{2,3}')} & \gate{R_z^{\ct}(\theta_{2,2}')} & \gate{R_x^{\ct}(\theta_{2,1}')} & \qw & \ctrl{-1} & \ctrl{1} & \gate{R_y^{\ct}(\frac{\pi}{2})} & \meter & \cw \\
			\Ket{0} & & \gate{R_y(\frac{\pi}{2})} & \ctrl{1} & \ctrl{-1} & \qw & \gate{R_x(\theta_{3,1})} & \gate{R_z(\theta_{3,2})} & \gate{R_x(\theta_{3,3})} & \qw & \gate{R_x^{\ct}(\theta_{3,3}')} & \gate{R_z^{\ct}(\theta_{3,2}')} & \gate{R_x^{\ct}(\theta_{3,1}')} & \qw & \ctrl{1} & \ctrl{-1} & \gate{R_y^{\ct}(\frac{\pi}{2})} & \meter & \cw \\
                & & & & & & & & & & & & & & & & & \\
			\qvdots & & \qvdots & \quad \qvdots & & & & \qvdots & & & & \qvdots & & & \quad \qvdots & & \qvdots & \qvdots \\
                & & & & & & & & & & & & & & & & & \\
			\Ket{0} & & \gate{R_y(\frac{\pi}{2})} & \qw & \ctrl{-1} & \qw & \gate{R_x(\theta_{N,1})} & \gate{R_z(\theta_{N,2})} & \gate{R_x(\theta_{N,3})} & \qw & \gate{R_x^{\ct}(\theta_{N,3}')} & \gate{R_z^{\ct}(\theta_{N,2}')} & \gate{R_x^{\ct}(\theta_{N,1}')} & \qw & \qw & \ctrl{-1} & \gate{R_y^{\ct}(\frac{\pi}{2})} & \meter & \cw \\
			& & \dstick{\qd V} & & & & & \dstick{D_{\bd{x}}} & & & & \dstick{D_{\bd{x}'}^{\ct}} & & & & & \dstick{V^{\ct}\qd}
			\gategroup{1}{3}{7}{5}{.6em}{--}
			\gategroup{1}{3}{7}{5}{1.2em}{_\}}
			\gategroup{1}{7}{7}{9}{.6em}{--}
			\gategroup{1}{7}{7}{9}{1.2em}{_\}}
			\gategroup{1}{11}{7}{13}{.6em}{--}
			\gategroup{1}{11}{7}{13}{1.2em}{_\}}
			\gategroup{1}{15}{7}{17}{.6em}{--}
			\gategroup{1}{15}{7}{17}{1.2em}{_\}}
		}
	}
	\vspace{1em}
	\caption{%
            The specific circuit of the quantum kernel proposed in~\cite{glick_covariant_2024} to classify cosets of $SU(2)^{\tp N}$\,.
		}
\label{fig:KernelCircuitChainGraph}
	\end{center}
\end{figure}
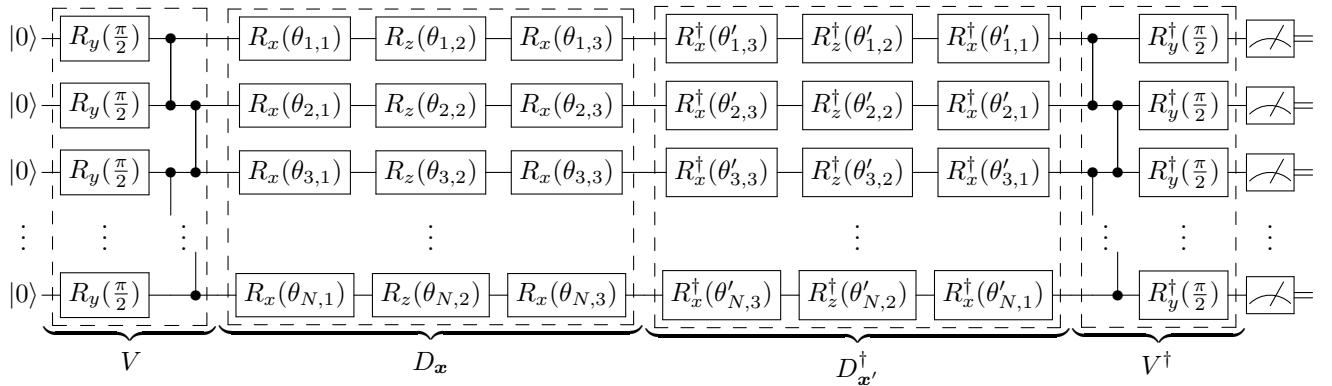

For this particular choice of graph-stabilizer group, the fiducial state is
\eqn{
    \Ket{\psi} = \tx{CZ}_{(1,2)}\tx{CZ}_{(2,3)}\tx{CZ}_{(3,4)} \dotsb \tx{CZ}_{(N-1,N)} \bigg(R_y\lb(\frac{\pi}{2}\rb)\Ket{0}\bigg)^{\tp N}
    \label{eq:RefState}
}
and the full quantum circuit for calculating the kernel value $\kappa(\bd{x},\bd{x}')$ in this case is given by Figure~\ref{fig:KernelCircuitChainGraph}.

Finally, we note that for ease of numerical simulations, we work with the generators of the group (Equation~\eqref{eq:ChainGenerators}) to form the set $S$, so that $n=N$, rather than the full group $S_{\rm graph}$, which has $2^N$ elements. As a consequence, our simulated data consists of the union of $m$ disjoint sets, $\mc{X} = \bigcup_{i=1}^{m} \bd{c}_jS$,
rather than $m$ cosets. To see that the sets are indeed disjoint, consider the full subgroup, $S_{\tx{graph}}$ and note that $S\subset S_{\tx{graph}}$. So, if $\bd{x}\in\bd{c}_iS$ then $\bd{x}\in\bd{c}_iS_{\tx{graph}}$. Since cosets are disjoint then $\bd{x}\notin\bd{c}_jS_{\tx{graph}}$ for all $j\ne i$, and therefore $\bd{x}\notin\bd{c}_jS$. Additionally, given the fiducial state Equation~\eqref{eq:RefState}, then 
\eqns{
    \bd{s}\Ket{\psi} = \Ket{\psi} \tx{ for all }\bd{s}\in S_{\tx{graph}} \imp \bd{s}\Ket{\psi} = \Ket{\psi} \tx{ for all }\bd{s}\in S.
}
Therefore, our simulated kernel will have its entries given by Equation~\eqref{eqn:alpha}.

\subsection{Properties of Operator Norms}
\label{appx:OpeartorNorm}

Here, we present some properties of the operator norm which we will use in calculations in later appendices.

Let $A: V\to W$ be a linear operator between normed spaces. The \textit{operator norm} of $A$ is (for $V\ne\{0\}$)
\begin{equation*}
	\opno{A} = \sup\lb\{\frac{\|Av\|}{\|v\|}:\ v\ne0 \tx{ and } v\in V\rb\}
\end{equation*}
where $\|v\|$ is the norm in $V$ and $\|Av\|$ is the norm in $W$ \cite{bhatia_matrix_1997-1}.

\noindent\textbf{Statement}: The first useful observation of the operator norm is that if $\opno{A-B_1}\le\epsilon$ and $\opno{A-B_2}\le\epsilon$ then
\eqn{
    \opno{B_2-B_1}\le2\epsilon.
    \label{eq:opnoDiff1}
}
\noindent\textbf{Proof}:
\alns{
    \opno{B_1-B_2} &= \opno{B_1-A+A-B_2} \nn\\
    &\le \opno{B_1-A} + \opno{A-B_2} \nn\\
    &\le \epsilon + \epsilon \nn\\
    &= 2\epsilon.
}

\noindent\textbf{Statement}: Next, notice that if $\opno{A_1-B_1}\le\epsilon$ and $\opno{A_2-B_2}\le\epsilon$ then
\eqn{
    \opno{A_1A_2-B_1B_2}\le\epsilon\min\big\{\opno{A_1}+\opno{B_2},\opno{A_2}+\opno{B_1}\big\}.
    \label{eq:opnoDiff2}
}
\textbf{Proof} \cite{watrous_theory_2018-1}:
\alns{
    \opno{A_1A_2-B_1B_2} &= \opno{A_1A_2-A_1B_2+A_1B_2-B_1B_2} \nn\\
    &\le \opno{A_1A_2-A_1B_2} + \opno{A_1B_2-B_1B_2} \nn\\
    &\le \opno{A_1}\opno{A_2-B_2} + \opno{B_2}\opno{A_1-B_1} \nn\\
    &\le \epsilon\big(\opno{A_1}+\opno{B_2}\big).
}
A similar argument can be used to show that
\eqns{
    \opno{A_1A_2-B_1B_2}\le\epsilon\big(\opno{A_2}+\opno{B_1}\big).
}
Combining these two bounds gives Equation~\eqref{eq:opnoDiff2}.

\noindent\textbf{Statement}: Finally, notice that $\opno{A-B}\le\epsilon$ then
\eqn{
    \opno{A}-\epsilon \le \opno{B} \le \opno{A}+\epsilon,
    \label{eq:opnoDiff3a}
}
and in particular if $A$ is unitary so $\opno{A}=1$, then
\eqn{
    1-\epsilon \le \opno{B} \le 1+\epsilon.
    \label{eq:opnoDiff3b}
}
\noindent\textbf{Proof}:
\alns{
    \epsilon \ge \opno{A-B} \ge \Big|\opno{A}-\opno{B}\Big| \quad (\tx{Reverse triangle inequality}).
}

\noindent\textbf{Statement}: Next, we write down three relationships between the operator norm and the overlap between states (vectors in an an inner product space of unit norm).
The first relationship is
\begin{equation}
    \opno{A}\ge\Abs{\Bk{\phi|A|\psi}}
    \label{eq:opnoIP1}
\end{equation}
for states $\Ket{\psi}$ and $\Ket{\phi}$.
\\\noindent\textbf{Proof}:
Recall that the induced norm of an inner product space is $\|v\| = \sqrt{\Bk{v|v}}$.
Therefore, we can bound
\alns{
	\opno{A} &\ge \frac{\|A\Ket{\psi}\|}{\|\Ket{\psi}\|} \nn\\
	&= \|A\Ket{\psi}\| \nn\\
	&= \|A\Ket{\psi}\|\ \|\Ket{\phi}\| \nn\\
	&\ge \Abs{\Bk{\psi|A|\phi}} \quad (\tx{Cauchy-Schwartz}).
}

\noindent\textbf{Statement}: The second relationship states that if $\opno{A-B}\le\delta$ then \eqn{
    \big|\Abs{\Bk{\phi|A|\psi}}-\Abs{\Bk{\phi|B|\psi}}\big|\le\delta.
    \label{eq:opnoIP2}
}
\noindent\textbf{Proof}:
\alns{
	\delta &\ge \opno{A-B} \nn\\
	&\ge \Abs{\Bk{\phi|(A-B)|\psi}} \nn\\
	&= \Big|\Bk{\phi|A|\psi}-\Bk{\phi|B|\psi}\Big| \nn\\
	&\ge \Big|\Abs{\Bk{\phi|A|\psi}}-\Abs{\Bk{\phi|B|\psi}}\Big| \quad (\tx{Reverse triangle inequality}).
	\label{eq:opnoIP2}
}

\noindent\textbf{Statement}: The third relationship states that if $\opno{U_1-U_2}\le\delta$ for unitary operators $U_1$ and $U_2$, then for any state $\Ket{\psi}$
\eqn{
    \Abs{\Bk{\psi|U_1^{\ct}U_2|\psi}}\ge 1-\delta^2/2.
    \label{eq:opnoIP3}
}
\noindent\textbf{Proof}:
\alns{
	\delta^2 &\ge \opno{U_1-U_2}^2 \nn\\
	&\ge \|(U_1-U_2)\Ket{\psi}\|^2 \nn\\
	&= \Big\langle(U_1-U_2)\psi \Big| (U_1-U_2)\psi\Big\rangle \nn\\
	&= \Bk{\psi|U_1^{\ct}U_1|\psi} - \Bk{\psi|U_1^{\ct}U_2|\psi} - \Bk{\psi|U_2^{\ct}U_1|\psi} + \Bk{\psi|U_2^{\ct}U_2|\psi} \nn\\
	&= 2\bigg(1-\Re\lb[\Bk{\psi|U_1^{\ct}U_2|\psi}\rb]\bigg) \nn\\
	&\ge 2\bigg(1-\Abs{\Bk{\psi|U_1^{\ct}U_2|\psi}}\bigg) \quad \tx{since } \Re[z]\le|z| \imp -\Re[z]\ge-|z|.
}
Rearranging gives Equation~\eqref{eq:opnoIP3}.

\subsection{Errors in the Fiducial State}
\label{appx:errorFiducial}
In the following we derive the bound discussed in Section~\ref{sec:results_noise_errorFiducial} of the main paper. For this, we assume that there is an error in the operator $V$ which takes the all zero state $\Ket{0^{\tp N}}$ to the fiducial state $\Ket{\psi}$ such that
\begin{equation*}
    	\opno{V-W}\le \epsilon.
\end{equation*}

If this is the only error, then each kernel element will take the form 
\begin{equation*}
   	\Abs{\Bk{0^{\tp N}|W_k^{\ct}D_{\bd{s}_a}^{\ct}D_{\bd{c}_i}^{\ct}D_{\bd{c}_j}D_{\bd{s}_b}W_{\ell}|0^{\tp N}}}^2 = \Abs{\Bk{0^{\tp N}|W_k^{\ct}UW_{\ell}|0^{\tp N}}}^2 
\end{equation*}
where $U=D_{\bd{s}_a}^{\ct}D_{\bd{c}_i}^{\ct}D_{\bd{c}_j}D_{\bd{s}_b}$. We also note that in general $W_k\ne W_{\ell}$, such as if the kernel is being implemented via the SWAP test \cite{thanasilp_exponential_2022}.

We can bound
\begin{align*}
	&\bigg|\Abs{\Bk{0^{\tp N}|V^{\ct}UV|0^{\tp N}}} - \Abs{\Bk{0^{\tp N}|W_k^{\ct}UW_{\ell}|0^{\tp N}}}\bigg| \nn\\
	&\quad \le \bigg|\Bk{0^{\tp N}|V^{\ct}UV|0^{\tp N}}-\Bk{0^{\tp N}|W_k^{\ct}UW_{\ell}|0^{\tp N}}\bigg| \quad (\tx{Reverse triangle inequality}) \nn\\
	&\quad = \Abs{\Bk{0^{\tp N}|\lb(V^{\ct}UV - W_k^{\ct}UW_{\ell}\rb)|0^{\tp N}}} \nn\\
	&\quad \le \opno{V^{\ct}UV - W_k^{\ct}UW_{\ell}} \quad \tx{by Equation~\eqref{eq:opnoIP1}} \nn\\
	&\quad \le \epsilon\lb(\opno{V^{\ct}U} + \opno{W_\ell}\rb) \quad \tx{by Equation~\eqref{eq:opnoDiff2}} \nn\\
    &\quad = \epsilon\lb(1 + \opno{W_\ell}\rb) \nn\\
    &\quad \le \epsilon\big(1+1+\epsilon) \quad \tx{by Equation~\eqref{eq:opnoDiff3b}} \nn\\
    &\quad = 2\epsilon+\epsilon^2\,.
\end{align*}

Therefore, in the case of $\bd{c}_i=\bd{c}_j$, where $\Abs{\Bk{0^{\tp N}|V^{\ct}UV|0^{\tp N}}}=1$, we have
\alns{
    	\Abs{\Bk{0^{\tp N}|W_k^{\ct}D_{\bd{s}_a}^{\ct}D_{\bd{c}_i}^{\ct}D_{\bd{c}_i}D_{\bd{s}_b}W_{\ell}|0^{\tp N}}} &\ge 1 - 2\epsilon - \epsilon^2 \nn\\
        \imp \Abs{\Bk{0^{\tp N}|W_k^{\ct}D_{\bd{s}_a}^{\ct}D_{\bd{c}_i}^{\ct}D_{\bd{c}_i}D_{\bd{s}_b}W_{\ell}|0^{\tp N}}}^2 &\ge 1 - 4\epsilon + 2\epsilon^2 + 4\epsilon^3 + \epsilon^4
}

and in the case of $\bd{c}_i\ne\bd{c}_j$, where $\Abs{\Bk{0^{\tp N}|V^{\ct}UV|0^{\tp N}}}=\sqrt{\alpha}$, we have
\aln{
	\sqrt{\alpha} - 2\epsilon - \epsilon^2 &\le \Abs{\Bk{0^{\tp N}|W_k^{\ct}D_{\bd{s}_a}^{\ct}D_{\bd{c}_i}^{\ct}D_{\bd{c}_j}D_{\bd{s}_b}W_{\ell}|0^{\tp N}}}  \le \sqrt{\alpha} + 2\epsilon + \epsilon^2 \nn\\
    \imp \alpha - 4\sqrt{\alpha}\epsilon + 2(2-\sqrt{\alpha})\epsilon^2 + 4\epsilon^3 + \epsilon^4 &\le \Abs{\Bk{0^{\tp N}|W_k^{\ct}D_{\bd{s}_a}^{\ct}D_{\bd{c}_i}^{\ct}D_{\bd{c}_j}D_{\bd{s}_b}W_{\ell}|0^{\tp N}}}^2  \le \alpha + 4\sqrt{\alpha}\epsilon + 2(2+\sqrt{\alpha})\epsilon^2 + 4\epsilon^3 + \epsilon^4.
    \label{eqn:PsiErrorDifferent}
}



Finally, notice if $\alpha\to 0$, then Equation~\eqref{eqn:PsiErrorDifferent} becomes
\begin{equation*}
    \Abs{\Bk{0^{\tp N}|W_k^{\ct}D_{\bd{s}_a}^{\ct}D_{\bd{c}_i}^{\ct}D_{\bd{c}_j}D_{\bd{s}_b}W_{\ell}|0^{\tp N}}} \le 2\epsilon + \epsilon^2 \imp \Abs{\Bk{0^{\tp N}|W_k^{\ct}D_{\bd{s}_a}^{\ct}D_{\bd{c}_i}^{\ct}D_{\bd{c}_j}D_{\bd{s}_b}W_{\ell}|0^{\tp N}}}^2 \le 4\epsilon^2 + 4\epsilon^3 + \epsilon^4
\end{equation*}
since an inner product must be non-negative.

\subsection{Simulation of errors in the fiducial state}
\label{appx:noisy1}
Here we discuss how to simulate the errors in the fiducial state presented in Section~\ref{sec:results_noise_errorFiducial} of the main paper.
For that, we present the details of the numerical simulation of the kernel matrix when there is an error in the creation of the fiducial state.

Recall, for our simulations, we take the operator $V$, which takes the all zero state, $\Ket{0}^{\tp N}$ to the fiducial state $\Ket{\psi}$ (given by Equation~\eqref{eq:RefState})to be
\eqns{
    V = \bigotimes_{j=1}^{N-1}\tx{CZ}_{(j,j+1)} \bigotimes_{j=1}^{N} R_y\lb(\frac{\pi}{2}\rb).
}

Consider the new operator
\eqn{
    W = \bigotimes_{j=1}^{N-1}\tx{CZ}_{(j,j+1)} \bigotimes_{j=1}^{N} R_y\lb(\frac{\pi}{2}-\theta_{j}\rb)
    \label{eq:W}
}
which is our choice of error in operator $V$.

The operator norm, see Appendix~\ref{appx:OpeartorNorm}, between the two operators is
\alns{
    \opno{V-W} &= \opno{\bigotimes_{j=1}^{N-1}\tx{CZ}_{(j,j+1)} \bigotimes_{j=1}^{N} R_y\lb(\frac{\pi}{2}\rb) - \bigotimes_{j=1}^{N-1}\tx{CZ}_{(j,j+1)} \bigotimes_{j=1}^{N} R_y\lb(\frac{\pi}{2}-\theta_j\rb)} \nn\\
    &=\opno{\bigotimes_{j=1}^{N-1}\tx{CZ}_{(j,j+1)}\lb(\bigotimes_{j=1}^{N} R_y\lb(\frac{\pi}{2}\rb)-\bigotimes_{j=1}^{N} R_y\lb(\frac{\pi}{2}-\theta_j\rb)\rb)} \nn\\
    &= \opno{\bigotimes_{j=1}^{N} R_y\lb(\frac{\pi}{2}\rb)-\bigotimes_{j=1}^{N} R_y\lb(\frac{\pi}{2}-\theta_j\rb)}.
}

This can be further simplified by using the fact that
\eqns{
    R_y\lb(\frac{\pi}{2}-\theta\rb) = R_y\lb(\frac{\pi}{2}\rb) R_y\big(-\theta\big)
}
so that
\alns{
    \opno{V-W} &= \opno{\bigotimes_{j=1}^{N} R_y\lb(\frac{\pi}{2}\rb)-\bigotimes_{j=1}^{N} R_y\lb(\frac{\pi}{2}\rb)R_y\big(-\theta_j\big)} \nn\\
    &= \opno{\bigotimes_{j=1}^{N} R_y\lb(\frac{\pi}{2}\rb)-\bigotimes_{j=1}^{N} R_y\lb(\frac{\pi}{2}\rb) \bigotimes_{j=1}^{N} R_y\big(-\theta_j\big)} \nn\\
    &= \opno{\bigotimes_{j=1}^{N} R_y\lb(\frac{\pi}{2}\rb) \lb(\id -\bigotimes_{j=1}^{N} R_y\big(-\theta_j\big)\rb)} \nn\\
    &= \opno{\id -\bigotimes_{j=1}^{N} R_y\big(-\theta_j\big)}.
}

Aside: If $A$ is normal $(A^{\ct}A=AA^{\ct})$, then $A = UDU^{\ct}$
where $U$ is a unitary matrix and $D$ is a diagonal matrix of the eigenvalues, $\lambda_j(A)$, of $A$ and
\eqns{
	\rho(A) = \opno{A}
}
where 
\eqns{
	\rho(A) = \max\{|\lambda_1(A)|, |\lambda_2(A)|, \dotsc, |\lambda_n(A)|\}
}
is the spectral radius of $A$. Additionally, for normal $A$, 
\eqns{
	\sigma_j(A) = |\lambda_j(A)|
}
where $\sigma(A)$ are the singular values of $A$.

Finally, notice that if $A$ is normal, then
\eqns{
	A-\id = UDU^{\ct} - \id = UDU^{\ct} - UU^{\ct} = U(D-\id)U^{\ct}
}
where
\alns{
	D-\id &= \pmx{
		\lambda_1(A) & 0 & \hdots & 0 \\
		0 & \lambda_2(A) & \hdots & 0 \\
		\vdots & \vdots & \ddots & \vdots \\
		0 & 0 & \hdots & \lambda_n(A)
	} - \pmx{
		1 & 0 & \hdots & 0 \\
		0 & 1 & \hdots & 0 \\
		\vdots & \vdots & \ddots & \vdots \\
		0 & 0 & \hdots & 1
	}
	 = \pmx{
		\lambda_1(A)-1 & 0 & \hdots & 0 \\
		0 & \lambda_2(A)-1 & \hdots & 0 \\
		\vdots & \vdots & \ddots & \vdots \\
		0 & 0 & \hdots & \lambda_n(A)-1
	}
}
and it is clear that $(A-\id)$ is normal with eigenvalues $\lambda_j(A-\id)=\lambda_j(A)-1$.
And so the square of the $j$th singular value of $A-\id$ is
\aln{
	\sigma_j(A-\id)^2 &= \Abs{\lambda_j(A-\id)}^2 \nn\\
	&= \Abs{\lambda_j(A)-1}^2 \nn\\
	&= \big(\lambda_j(A) - 1\big)\big(\lambda_j^*(A) - 1\big) \nn\\
	&= 1 + \Abs{\lambda_j(A)}^2 - 2\Re\big[\lambda_j(A)\big]
    \label{eq:UsefulSV}
}
and if $A$ is unitary then
\eqns{
    \sigma_j(A-\id)^2 = 2 - 2\Re\big[\lambda_j(A)\big].
}

Since the operator norm, $\opno{V-W}$ is of the form $\opno{U-\id}$ for a unitary operator, $U$, we can use Equation~\eqref{eq:UsefulSV} to bound the maximum possible value that each $\theta_j$ can take to keep $\opno{V-W}\le\epsilon$.

The two eigenvalues of $R_y(-\theta_j)$ are
\eqns{
    \lambda_{\pm}\big(R_y(-\theta_j)\big) = \cos\lb(\frac{\theta_j}{2}\rb) \pm \iu\sin\lb(\frac{\theta_j}{2}\rb)
}
which we note are the same pair of eigenvalues of $R_y(+\theta_j)$, so for simplicity, we will consider the eigenvalues of $R_y\big(\Abs{\theta_j}\big)$. For small angles, $\theta_j= x_j\delta$ with $\delta\ll1$, the eigenvalues can be approximated as
\eqns{
    \lambda_{k_j}\Big(R_y\big(\Abs{\theta_j}\big)\Big) = 1 + (-1)^{k_j} \frac{\iu \Abs{x_j}}{2} \delta - \frac{x_j^2}{8}\delta^2 + \Or(\delta^3)
}
where we now encode the sign $(\pm)$ into the value of $k_j\in\{0,1\}$.

Now, we use the property that if $A$ has $m$ eigenvalues $\lambda_j(A)$ and if $B$ has $n$ eigenvalues $\lambda_k(B)$ then $A\tp B$ has $mn$ eigenvalues
\eqns{
	\lambda_{(j,k)}(A\tp B) = \lambda_j(A)\lambda_k(B)
}
to calculate the the square of the $\bd{k}=(k_1,k_2,\dotsc,k_N)$-th singular value (unordered) of $\bigotimes_{j=1}^{N}R_{y}\big(\Abs{\theta_j}\big) - \id$  
\alns{
	&\sigma_{\bd{k}}^2\lb(\bigotimes_{j=1}^{N}R_y\big(|\theta_j|\big) - \id \rb) \nn\\
	&\qd = 2\lb\{1 - \Re\lb[\prod_{j=1}^{N}\lambda_{k_j}\Big(R_y\big(|\theta_j|\big)\Big)\rb]\rb\} \nn\\
	&\qd = 2\lb\{1 - \Re\lb[1 +\frac{\iu\delta}{2} \sum_{j=1}^{N} (-1)^{k_j}|x_j| - \frac{\delta^2}{8} \sum_{j=1}^{N} x_j^2 - \frac{\delta^2}{4}\sum_{i=1}^{N-1}\sum_{j=i+1}^{N}(-1)^{k_i+k_j}|x_i||x_j|\rb]\rb\} + \Or(\delta^3) \nn\\
	&\qd = 2\lb(\frac{\delta^2}{8}\sum_{j=1}^{N} x_j^2 + \frac{\delta^2}{4} \sum_{i=1}^{N-1}\sum_{j=i+1}^{N}(-1)^{k_i+k_j}|x_i||x_j|\rb) + \Or(\delta^3).
}

This is maximized when the parity of $k_i$ and $k_j$ is the same for all $i,j$ (take either $\bd{k}=(0,0,\dotsc,0)$ or ${\bd{k}=(1,1,\dotsc,1)}$), so
\alns{
	\sigma_{\tx{max}}^2\lb(\bigotimes_{j=1}^{N}R_y\big(|\theta_j|\big) - \id \rb) = \frac{\delta^2}{4} \lb(\sum_{j=1}^{N} x_j^2 + 2\sum_{i=1}^{N-1}\sum_{j=i+1}^{N}|x_i||x_j|\rb) + \Or(\delta^3).
}

Now take $y=\max\big\{|x_1|,|x_2|,\dotsc,|x_N|\big\}$, so to lowest order in $\delta$.
\alns{
	\sigma_{\tx{max}}^2\lb(\bigotimes_{j=1}^{N}R_y\big(|\theta_j|\big) - \id \rb) &\le \frac{\delta^2}{4} \lb(\sum_{j=1}^{N} y^2 + 2\sum_{i=1}^{N-1}\sum_{j=i+1}^{N}y^2\rb) \nn\\
	&= \frac{(y\delta)^2}{4} \lb(N+2\frac{N(N-1)}{2}\rb) \nn\\
	%
	%
	&= \frac{N^2}{4} (y\delta)^2.
}

Therefore, if
\eqns{
	\Abs{\theta_j} \le y\delta = \frac{2\epsilon}{N}
}
for all $j\in\{1,\dotsc,N\}$ then
\alns{
	\opno{V-W}^2 = \opno{\bigotimes_{j=1}^{N} R_y\big(-\theta_y\big) - \id}^2 = \sigma_{\tx{max}}^2\lb(\bigotimes_{j=1}^{N}R_y\big(|\theta_j|\big) - \id \rb) \le \frac{N^2}{4} (y\delta)^2 = \frac{N^2}{4} \lb(\frac{2\epsilon}{N}\rb)^2 = \epsilon^2.
}

\begin{figure}[ht]
	\begin{center}
	\mbox{
		\Qcircuit @C=1em @R=0.7em {
	 		\Ket{0} & & \multigate{3}{W_1} & \multigate{3}{D_{\bd{x}}} & \qw & \multigate{3}{D_{\bd{x}'}^{\ct}} & \multigate{3}{W_2^{\ct}} & \meter & \cw \\
			\Ket{0} & & \ghost{W_1} & \ghost{D_{\bd{x}}} & \qw & \ghost{D_{\bd{x}'}^{\ct}} & \ghost{W_2^{\ct}} & \meter & \cw \\
			\qvdots & & \nghost{W_1}  & \nghost{D_{\bd{x}}} & & \nghost{D_{\bd{x}'}^{\ct}} & \nghost{W_2^{\ct}} & \qvdots \\
			\Ket{0} & & \ghost{W_1} & \ghost{D_{\bd{x}}} & \qw & \ghost{D_{\bd{x}'}^{\ct}} & \ghost{W_2^{\ct}} & \meter & \cw 
		}
	}
	\vspace{1em}
	\caption{Quantum circuit for calculating the kernel elements when there are errors in the creation of the fiducial state.
		}
        \label{fig:KernelCircuitError1}
	\end{center}
\end{figure}
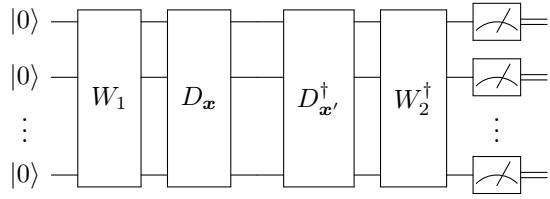

For each simulation of the kernel matrix, we select the $N$ values of $\theta_j$ from a uniform distribution $[-2\epsilon/N,2\epsilon/N]$ to create $W_1$ (equation~\eqref{eq:W}) and another $N$ values of $\theta_j$ to create $W_2$. We simulate the quantum circuit shown in Figure~\ref{fig:KernelCircuitError1}, calculate the variance in the non-trivial kernel values, and repeat 100 times, each with new values of $\theta_j$ for $W_1$ and $W_2$.

\subsection{Error in the unitary representation of the elements of the group}
\label{appx:unitaryRep}
We now prove the bound stated in Section~\ref{sec:results_noise_unitary} of the main paper. We assume that there is an error in the unitary representation of the group elements so that for each $\bd{x}_{a,i}$ the operator $A_{a,i}$ is used rather than $D_{\bd{x}_{a,i}}$ such that
\eqns{
    \opno{D_{\bd{x}_{a,i}}-A_{a,i}} \le \epsilon.
}

If this is the only error, then each kernel element will take the form 
\eqns{
    \Abs{\Bk{0^{\tp N}|V^{\ct}A_{a,i}^{\ct}A_{b,j}V|0^{\tp N}}}^2 = \Abs{\Bk{\psi|A_{a,i}^{\ct}A_{b,j}|\psi}}^2.
}

Following the same steps as Appendix~\ref{appx:errorFiducial}, we can bound
\eqns{
    \bigg| \Abs{\Bk{\psi|D_{\bd{x}_{a,i}}^{\ct}D_{\bd{x}_{b,j}}|\psi}} - \Abs{\Bk{\psi|A_{a,i}^{\ct}A_{b,j}|\psi}}\bigg| \le 2\epsilon + \epsilon^2
}
so in the case of $i=j$, where $\Abs{\Bk{\psi|D_{\bd{x}_{a,i}}^{\ct}D_{\bd{x}_{b,i}}|\psi}}=1$, we have
\eqns{
    \Abs{\Bk{\psi|A_{a,i}^{\ct}A_{b,i}|\psi}}^2 \ge 1 - 4\epsilon + 2\epsilon^2 + 4\epsilon^3 + \epsilon^4
}

and in the case of $i\ne j$, where $\Abs{\Bk{\psi|D_{\bd{x}_{a,i}}^{\ct}D_{\bd{x}_{b,j}}|\psi}}=\sqrt{\alpha}$, we have
\eqns{
    \alpha - 4\sqrt{\alpha}\epsilon + 2(2-\sqrt{\alpha})\epsilon^2 + 4\epsilon^3 + \epsilon^4 \le \Abs{\Bk{\psi|A_{a,i}^{\ct}A_{b,j}|\psi}}^2  \le \alpha + 4\sqrt{\alpha}\epsilon + 2(2+\sqrt{\alpha})\epsilon^2 + 4\epsilon^3 + \epsilon^4.
}



\subsection{Error in the selection of the group elements }
\label{appx:selectionError}
In what follows, we present the proof of the bound from Section~\ref{sec:results_noise_selectionError}.
As presented in \cite{glick_covariant_2024}, we assume that for each $1\le a\le n$ and $1 \le i\le m$ there is a random perturbation $\bd{e}_{a,i}\in\mc{G}$ such that
\begin{equation*}
    d(\bd{e}_{a,i},\id) \ce \opno{D_{\bd{e}_{a,i}}-\id} < \epsilon
\end{equation*}
and for each $\bd{q}_{a,i}\in\bigcup_{j}C_j$ we have an $\bd{x}_{a,i}\in\mc{X}$ such that $\bd{x}_{a,i}=\bd{e}_{a,i}\bd{q}_{a,i}$. Since $D$ is a unitary representation of the group $\mc{G}$, we have
\begin{equation*}
    D_{\bd{x}_{a,i}} = D_{\bd{e}_{a,i}}D_{\bd{q}_{a,i}} = D_{\bd{e}_{a,i}}D_{\bd{c}_{i}}D_{\bd{s}_a}.
\end{equation*}
So, assuming no more errors, each kernel element will take the form 
\begin{align*}
    \Abs{\Bk{0^{\tp N}|V^{\ct}D_{\bd{s}_a}^{\ct}D_{\bd{c}_i}^{\ct}D_{\bd{e}_{a,i}}^{\ct}D_{\bd{e}_{b,j}}D_{\bd{c}_{j}}D_{\bd{s}_b}V|0^{\tp N}}}^2 &= \Abs{\Bk{\psi|D_{\bd{s}_a}^{\ct}D_{\bd{c}_i}^{\ct}D_{\bd{e}_{a,i}}^{\ct}D_{\bd{e}_{b,j}}D_{\bd{c}_{j}}D_{\bd{s}_b}|\psi}}^2 \nn\\
    &= \Abs{\Bk{\psi|D_{\bd{c}_i}^{\ct}D_{\bd{e}_{a,i}}^{\ct}D_{\bd{e}_{b,j}}D_{\bd{c}_{j}}|\psi}}^2 \nn\\
    &= \Abs{\Bk{\varphi_i|D_{\bd{e}_{a,i}}^{\ct}D_{\bd{e}_{b,j}}|\varphi_j}}^2
\end{align*}
where we define $\Ket{\varphi_j}\ce D_{\bd{c}_j}\Ket{\psi}$.




In the case of $\bd{c}_i=\bd{c}_j$ where $\Abs{\Bk{\psi|D_{\bd{s}_a}^{\ct}D_{\bd{c}_i}^{\ct}D_{\bd{c}_i}D_{\bd{s}_b}|\psi}} = \Abs{\Bk{\varphi_i|\varphi_i}}=1$,
we first notice that 
\eqns{
    \opno{D_{\bd{e}_{a,i}}-D_{\bd{e}_{b,j}}} \le 2\epsilon
}
which follows from Equation~\eqref{eq:opnoDiff1}.
Then we can use Equation~\eqref{eq:opnoIP3} to bound
\alns{
    \Abs{\Bk{\varphi_i|D_{\bd{e}_{a,i}}^{\ct}D_{\bd{e}_{b,i}}|\varphi_i}} &\ge 1-\frac{(2\epsilon)^2}{2} = 1-\frac{\epsilon^2}{2} \nn\\
    \imp \Abs{\Bk{\varphi_i|D_{\bd{e}_{a,i}}^{\ct}D_{\bd{e}_{b,i}}|\varphi_i}}^2 &\ge 1-\epsilon^2+\frac{\epsilon^4}{4}.
}

In the case of $\bd{c}_i\ne\bd{c}_j$ where $\Abs{\Bk{\psi|D_{\bd{s}_a}^{\ct}D_{\bd{c}_i}^{\ct}D_{\bd{c}_j}D_{\bd{s}_b}|\psi}} = \Abs{\Bk{\varphi_i|\varphi_j}}=\sqrt{\alpha}$,
we first notice that (since the operator norm, see Appendix~\ref{appx:OpeartorNorm}, is an unitarily invariant norm)
\eqns{
    2\epsilon \ge \opno{D_{\bd{e}_{a,i}}-D_{\bd{e}_{b,j}}} = \opno{D_{\bd{e}_{a,i}}^{\ct}D_{\bd{e}_{a,i}}-D_{\bd{e}_{a,i}}^{\ct}D_{\bd{e}_{b,j}}} = \opno{\id -D_{\bd{e}_{a,i}}^{\ct}D_{\bd{e}_{b,j}}}
}



so we can now bound
\begin{align}
	\bigg|\Abs{\Bk{\varphi_i|D_{\bd{e}_{a,i}}^{\ct}D_{\bd{e}_{b,j}}|\varphi_j}} - \Abs{\Bk{\varphi_i|\varphi_j}}\bigg| &\le \bigg|\Bk{\varphi_i|D_{\bd{e}_{a,i}}^{\ct}D_{\bd{e}_{b,j}}|\varphi_j} - \Bk{\varphi_i|\varphi_j}\bigg| \quad (\text{Reverse triangle inequality}) \nn\\
	&= \Abs{\Bk{\varphi_i|\lb(D_{\bd{e}_{a,i}}^{\ct}D_{\bd{e}_{b,j}}-\id\rb)|\varphi_j}} \nn\\
	&\le \opno{D_{\bd{e}_{a,i}}^{\ct}D_{\bd{e}_{b,j}}-\id} \quad \tx{by Equation~\eqref{eq:opnoIP1}}\nn\\
	&\le 2\epsilon. \label{eqn:fact4}
\end{align}

Therefore, we have (for $i\ne j$)
\begin{align*}
    \sqrt{\alpha} - 2\epsilon &\le \Abs{\Bk{\psi|D_{\bd{s}_a}^{\ct}D_{\bd{c}_i}^{\ct}D_{\bd{e}_{a,i}}^{\ct}D_{\bd{e}_{b,j}}D_{\bd{c}_{j}}D_{\bd{s}_b}|\psi}}  \le \sqrt{\alpha} + 2\epsilon \nn\\
    \imp \alpha - 4\sqrt{\alpha}\epsilon + 4\epsilon^2 &\le \Abs{\Bk{\psi|D_{\bd{s}_a}^{\ct}D_{\bd{c}_i}^{\ct}D_{\bd{e}_{a,i}}^{\ct}D_{\bd{e}_{b,j}}D_{\bd{c}_{j}}D_{\bd{s}_b}|\psi}}^2  \le \alpha + 4\sqrt{\alpha}\epsilon + 4\epsilon^2.
\end{align*}




\subsection{Simulations of errors in the selection of the group elements} \label{appx:noisy}

In this section we discuss the details of how we simulated errors in the selection of the group elements, which we presented in Section~\ref{sec:results_noise_numerics}.

This type of error assumes a random perturbation, $\bd{e}\in\mc{G}$, on each $\bd{x}\in\mc{X}$ such that 
\begin{equation*}
    d(\bd{e},\id) \ce \opno{D_{\bd{e}}-\id} < \epsilon.
\end{equation*}

For our choice of group, $\mc{G}=\tx{SU}(2)^{\tp N}$, each perturbation can be represented as
\eqn{
    D_{\bd{e}} = \bigotimes_{j=1}^{N} R_x(\theta_{j,1})R_z(\theta_{j,2})R_x(\theta_{j,3}).
    \label{eq:De}
}

Similar to our simulation of errors in the fiducial state, we again use Equation~\eqref{eq:UsefulSV} to bound the maximum possible value that each $\theta_{j,i}$ can take to keep $\opno{D_{\bd{e}}-\id}\le\epsilon$.

The eigenvalues of $R_X(\theta_{j,1})R_Z(\theta_{j,2})R_X(\theta_{j,3})$ can easily be found to be
\eqns{
	\lambda_{\pm}\big(R_x(\theta_{j,1})R_z(\theta_{j,2})R_x(\theta_{j,3})\big)  = \cos\lb(\frac{\theta_{j,2}}{2}\rb)\cos\lb(\frac{\theta_{j,1}+\theta_{j,3}}{2}\rb) \pm \iu \sqrt{1 - \cos^2\lb(\frac{\theta_{j,2}}{2}\rb)\cos^2\lb(\frac{\theta_{j,1}+\theta_{j,3}}{2}\rb)}.
}

For small angles, $\theta_{j,i}= x_{j,i}\delta$ with $\delta\ll1$, the eigenvalues can be approximated as
\alns{
	&\lambda_{k_j}\big(R_x(x_{1,j}\delta)R_z(x_{j,2}\delta)R_x(x_{j,3}\delta)\big) \nn\\
	&\qd = 1 + (-1)^{k_j}\ \iu \sqrt{\frac{x_{j,2}^2}{4}+\frac{(x_{j,1}+x_{j,3})^2}{4}} \delta - \frac{1}{2}\lb(\frac{x_{j,2}^2}{4}+\frac{(x_{j,1}+x_{j,3})^2}{4}\rb)\delta^2 + \Or(\delta^3) \nn\\
	&\qd = 1 + (-1)^{k_j}\ \frac{\iu}{2} \sqrt{X_j} \delta - \frac{1}{8} X_j\delta^2 + \Or(\delta^3)
}
where the sign $(\pm)$ is represented as the value of $k_j\in\{0,1\}$ and we define
\eqns{
	X_j \ce x_{j,2}^2 + (x_{j,1}+x_{j,3})^2 \ge 0.
}

\begin{figure}[t!]
	\begin{center}
	\mbox{
		\Qcircuit @C=1em @R=0.7em {
	 		\Ket{0} & & \multigate{3}{V} & \multigate{3}{D_{\bd{x}}} & \multigate{3}{D_{\bd{e}}} & \qw & \multigate{3}{D_{\bd{e}'}^{\ct}} & \multigate{3}{D_{\bd{x}'}^{\ct}} & \multigate{3}{V^{\ct}} & \meter & \cw \\
			\Ket{0} & & \ghost{V} & \ghost{D_{\bd{x}}} & \ghost{D_{\bd{e}}} & \qw & \ghost{D_{\bd{e}'}^{\ct}} & \ghost{D_{\bd{x}'}^{\ct}} & \ghost{V^{\ct}} & \meter & \cw \\
			\qvdots & & \nghost{V}  & \nghost{D_{\bd{x}}} & \nghost{D_{\bd{e}}} & & \nghost{D_{\bd{e}'}^{\ct}} & \nghost{D_{\bd{x}'}^{\ct}} & \nghost{V^{\ct}} & \qvdots \\
			\Ket{0} & & \ghost{V} & \ghost{D_{\bd{x}}} & \ghost{D_{\bd{e}}} & \qw & \ghost{D_{\bd{e}'}^{\ct}} & \ghost{D_{\bd{x}'}^{\ct}} & \ghost{V^{\ct}} & \meter & \cw 
		}
	}
	\vspace{1em}
	\caption{Quantum circuit for calculating the kernel elements when there are errors in the selection of group elements.
		}
                \label{fig:KernelCircuitError2}
	\end{center}
\end{figure}
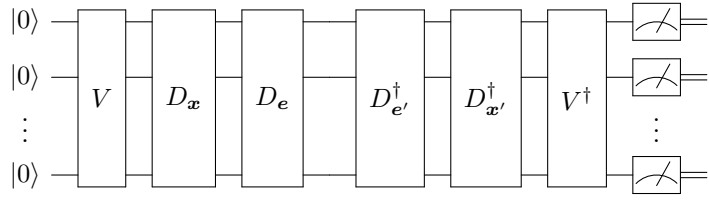

The square of the $\bd{k}=(k_1,k_2,\dotsc,k_N)$-th singular value (unordered) of $\lb(\bigotimes_{j=1}^{N}R_{x}(\theta_{j,1})R_z(\theta_{j,2})R_x(\theta_{j,3})\rb) - \id$ is
\alns{
	&\sigma_{\bd{k}}^2\lb[\lb(\bigotimes_{j=1}^{N} R_x(\theta_{j,1})R_z(\theta_{j,2})R_x(\theta_{j,3})\rb)-\id\rb] \nn\\
    &\qd = 2\lb\{1 - \Re\lb[\prod_{j=1}^{N}\lambda_{k_j}\big(R_x(\theta_{j,1})R_z(\theta_{j,2})R_x(\theta_{j,3})\big)\rb]\rb\} \nn\\
    &\qd = 2\Bigg(\frac{\delta^2}{8}\sum_{j=1}^{N} X_j + \frac{\delta^2}{4}\sum_{i=1}^{N-1}\sum_{j=i+1}^{n} (-1)^{k_i+k_j}\sqrt{X_iX_j}\Bigg) + \Or(\delta^3).
}

This is maximized when the parity of $k_i$ and $k_j$ is the same for all $i,j$ (take either $\bd{k}=(0,0,\dotsc,0)$ or ${\bd{k}=(1,1,\dotsc,1)}$), so
\alns{
	\sigma_{\tx{max}}^2\lb[\lb(\bigotimes_{j=1}^{N} R_x(\theta_{j,1})R_z(\theta_{j,2})R_x(\theta_{j,3})\rb)-\id\rb] = \frac{\delta^2}{4}\sum_{j=1}^{N} X_j + \frac{\delta^2}{2}\sum_{i=1}^{N-1}\sum_{j=i+1}^{N} \sqrt{X_iX_j} + \Or(\delta^3).
}

Now take $y=\max\big\{|x_{1,1}|,|x_{1,2}|,|x_{1,3}|,|x_{2,1}|, |x_{2,2}|,|x_{2,3}|,\dotsc,|x_{N,1}|,|x_{N,2}|,|x_{N,3}|\big\}$, so 
\eqns{
    X_j \le (y+y)^2+y^2 = 5y^2
}
and to lowest order in $\delta$
\alns{
    \sigma_{\tx{max}}^2\lb[\lb(\bigotimes_{j=1}^{N} R_x(\theta_{j,1})R_z(\theta_{j,2})R_x(\theta_{j,3})\rb)-\id\rb] &\le \frac{\delta^2}{4}\sum_{j=1}^{N} 5y^2 + \frac{\delta^2}{2}\sum_{i=1}^{N-1}\sum_{j=i+1}^{N} 5y^2 \nn\\
    &= \frac{5(y\delta)^2}{4}\lb(N+2\frac{N(N-1)}{2}\rb) \nn\\
    &= \frac{5N^2}{4}(y\delta)^2.
}
Therefore, if 
\eqns{
    |\theta_{j,i}| \le y\delta = \frac{2\epsilon}{\sqrt{5}N}
}
then
\alns{
    \opno{D_{\bd{e}}-\id}^2 &= \opno{\lb(\bigotimes_{j=1}^{N} R_x(\theta_{j,1})R_z(\theta_{j,2})R_x(\theta_{j,3})\rb)-\id} \nn\\
    &= \sigma_{\tx{max}}^2\lb[\lb(\bigotimes_{j=1}^{N} R_x(\theta_{j,1})R_z(\theta_{j,2})R_x(\theta_{j,3})\rb)-\id\rb] \nn\\
    &\le \frac{5N^2}{4}\lb(\frac{2\epsilon}{\sqrt{5}N}\rb)^2 \nn\\
    &= \epsilon^2.
}

For each $\bd{x}\in\mc{X}$ we select $3N$ values of $\theta_{j,i}$ from a uniform distribution of $\lb[-\frac{2\epsilon}{\sqrt{5}N}, \frac{2\epsilon}{\sqrt{5}N}\rb]$ to create $D_{\bd{e}}$ (equation~\eqref{eq:De}). We then simulate the quantum circuit shown in Figure~\ref{fig:KernelCircuitError2} for each pair $(\bd{x},\bd{x}')$ in the test data and the associated $(\bd{e},\bd{e}')$ and calculate the variance in the non-trivial kernel matrix. This is repeated 100 times, with new values of $\theta_{j,i}$ for each $D_{\bd{e}}$.


\end{document}